\documentclass[a4paper]{article}

\usepackage[english]{babel}
\usepackage[utf8x]{inputenc}
\usepackage[T1]{fontenc}
\usepackage{natbib}
\usepackage{array}
\usepackage{graphicx}
\usepackage{comment}
\usepackage{microtype}

\usepackage[a4paper,top=3cm,bottom=2cm,left=3cm,right=3cm,marginparwidth=1.75cm]{geometry}

\usepackage{amsmath}
\usepackage{graphicx}
\usepackage[colorinlistoftodos]{todonotes}
\usepackage[colorlinks=true, allcolors=blue]{hyperref}
\usepackage{authblk}
\usepackage{hyperref}
\usepackage[T1]{fontenc}

\title{Machine learning as a model for cultural learning: \\ Teaching an algorithm what it means to be fat}

\author[1]{Alina Arseniev-Koehler}
\author[2]{Jacob G. Foster} 
\affil[ ]{Department of Sociology, UCLA}
\affil[1]{\url{arsena@g.ucla.edu}}
\affil[2]{\url{foster@soc.ucla.edu}}

\date{}

\begin{document}
\maketitle

\begin{abstract} 
As we navigate our cultural environment, we learn cultural biases, like those around gender, social class, health, and body weight. It is unclear, however, exactly how public culture becomes private culture. In this paper, we provide a theoretical account of such cultural learning. We propose that neural word embeddings provide a parsimonious and cognitively plausible model of the representations learned from natural language. Using neural word embeddings, we extract cultural schemata about body weight from \textit{New York Times} articles. We identify several cultural schemata that link obesity to gender, immorality, poor health, and low socioeconomic class. Such schemata may be subtly but pervasively activated in public culture; thus, language can chronically reproduce biases. Our findings reinforce ongoing concerns that machine learning can also encode, and reproduce, harmful human biases. 
\end{abstract}

\vspace{5mm}
%

As we navigate our cultural environment, we learn to ascribe pejorative meanings to seemingly arbitrary differences. In the case of body weight, for example, media portrayals of fat and thin bodies may drive widespread idealization of thinness and fear of fat in the U.S \citep{grabe2008role, lopez2010influence, saguy2008fat, ravary2019shaping, ata2010weight}. While there is little doubt that public culture, like media, is a powerful source for cognitive socialization, it remains unclear exactly how an individual decodes and internalizes meanings in the cultural environment \citep{bail2014cultural}. Nor is it obvious how cultural meanings are encoded and reconstructed as they travel from public life to private thoughts \citep{lizardo2017improving,foster2018culture}. 

What are the cognitive and material processes that underlie the social construction of shared meanings? And how do cultural objects, like the text in news media, help to produce shared meanings? Our paper begins to tackle this gap between public and private culture, focusing on how we internalize public constructions of obesity. First, we suggest that schema learning is the core mechanism by which public culture becomes private culture. Second, we suggest that neural word embeddings, a language model inspired by human cognition, provide a parsimonious model for this cultural learning. Ultimately, we provide a cognitively plausible, computational model for learning private culture from public culture.  

Schematic processing is a key mechanism for cognitive socialization. We use the term schema (plural schemata) to denote the basic computational ``building blocks" of human knowledge and information processing \citep{rumelhart1980schemata,dimaggio1997culture}. Although \citet{rumelhart1980schemata} describes a schema as ``a data structure for representing the generic concepts stored in memory," he emphasizes that they are very much \textit{processes}, rather than static objects. Schemata are used to do things: to make sense of a scene, or to interpret a story. Materially implemented as patterns of neural connectivity in individual human brains \citep{wood2018schemas}, schemata are patterned ways of processing stimuli (such as a news article about body weight). On this account, any individual meaning-making necessarily involves creating, activating, reinforcing, or tuning schemata. This schema-driven information processing “occur[s] without our conscious awareness, and [it] underlie[s] the way we attach meaning to things” \citep{shepherd2011cultural}. Hence it is an \textit{implicit} form of memory, and is invoked automatically rather than deliberately \citep{schacter1987implicit}. Schemata are valuable cognitive tools because they allow us to fill in the blanks about the world; they are a "source of predictions about unobserved events" \citep{rumelhart1980schemata}.

Schemata are often cultural; whenever they are "shared regularit[ies] in the organization of experience acquired through social life" \citep{foster2018culture} we refer to them as \textit{cultural} schemata. Schemata of fat provide a prominent example. Indeed, overweight individuals are commonly perceived as lazy, immoral, unintelligent, and unsuccessful \citep{puhl2010obesity, schwartz2006influence};  nearly half (46\%) of Americans would be willing to give up at least 1 year of life rather than be obese \citep{schwartz2006influence}. News and media constructions of body weight are considered key influences for widespread internalized body dissatisfaction, eating disorders, anti-fat prejudice, and the widespread, even normalized, discrimination on the basis of body weight in the U.S. (e.g., \cite{saguy2008fat,grabe2008role, lopez2010influence, ravary2019shaping, lin2009relationship, ata2010weight, latner2007childhood,campos2005epidemiology}). However, it remains unclear exactly how public meanings of body weight are internalized. In this paper, we explicitly model the way that cultural schemata of obesity can be learned from, and activated by, cultural objects. Here, we focus on how people learn the cultural connotations of words from text. We draw on a contemporary machine-learning approach to learning and representing word meanings in natural language called word2vec \citep{mikolov2013distributed}. 

Word2vec uses an artificial neural network (ANN) to learn semantically rich representations of words. Its representational scheme and algorithmic structure are inspired (in broad strokes) by the architecture of the human brain and its learning processes \citep{miller1995assessment,mitchell2008predicting}. The overall computational strategy emerges from a long tradition  in cognitive science \citep{gunther2019vector} that models semantic memory as high-dimensional vectors learned from the way meaningful units are jointly distributed across experience \citep{jones2015models}. While an abundance of empirical work already demonstrates that the substantive meanings in word2vec and other embeddings powerfully correspond to private meaning (e.g., \cite{caliskan2017semantics, bolukbasi2016quantifying, kozlowski2019geometry, grand2018semantic, jones2019stereotypical, joseph2020word, garg2018word, dehghani2017decoding, vodrahalli2018mapping}), we argue that neural embeddings, like word2vec, are also cognitively plausible models for \textit{learning} private meaning.  As we will show, word2vec allows us to model the learning of schemata at multiple levels and at large-scale--while responding to older calls to integrate cultural sociology with cognition \citep{dimaggio1997culture, strauss1997cognitive} and more recent calls to integrate culture and cognition research with computational thinking \citep{foster2018culture}. 

Previous literature on obesity in public culture has focused on the framing of body weight: how cultural objects like news reporting guide the interpretation of obesity as gendered, immoral, unhealthy, and low-class (e.g., \cite{boero2007all,saguy2008fat, saguy2014reporting}). Meanwhile, a large body of work also documents our private meanings of body fat, suggesting that these are deeply influence by public culture like media (e.g., \cite{nosek2007pervasiveness, ravary2019shaping, carels2013stereotypical, teachman2003demonstrations, teachman2001implicit, hinman2015stereotypical, schwartz2006influence, schupp2011implicit,azevedo2014weighing, lieberman2012disgust}). We aim to close the theoretical circle implicit in this scholarship, by building computational models of the schemata learned from and invoked by the frames in news reporting.

After reviewing the rich literature on the social construction of meanings around body weight in the U.S., we provide a computationally inflected account of how schemata and frames enable the construction of meanings -- such as meanings attached to body weight. We build up to the key idea of \textit{distributed representation}, which allows us to think more rigorously about schemata in a neural-network architecture \citep{rumelhart1986sequential}. We then detail the workings of word2vec and its ANN architecture, highlighting the similarities to human cognition. We argue that just as topic modeling is a theoretically motivated method to study frames at scale \citep{dimaggio2013exploiting}, neural word embeddings offer a theoretically motivated method to study schemata and their acquisition. We then describe our method for learning schemata from text using word2vec, drawing out schemata at multiple levels (i.e., individual word schemata and higher-level schemata for concepts like gender or moralization). Finally, we apply our methods to quantify various schemata used to process information about body weight and health in news discourse. 

\vspace{5mm}
\begin{center}
\item{\section*{SOCIAL CONSTRUCTION OF BODY WEIGHT IN THE U.S.}}
\end{center}

In the United States, body fat is typically viewed as undesirable, especially for women. Indeed, most (64\%) of Americans want to lose weight \citep{yaemsiri2011perceived}.  Body fat is  not only viewed as unattractive, but also as unhealthy and within an individual's control. This makes body fat a target of moral as well as aesthetic judgment. These meanings are not inherent characteristics of fatness itself -- from a world-historical perspective, contemporary Western societies are anomalous in viewing fatness as undesirable and thinness as idealized \citep{ernsberger2009does}. Rather, these meanings are socially produced in everyday interaction and in institutions like the government, medicine, and news \citep{campos2005epidemiology, saguy2008fat}. Early campaigning against fat in the U.S. began at the turn of the 20th century (1890-1910) \citep{stearns2002fat}: pejorative meanings of body weight have become pervasive in public culture such as news, social media, and television (e.g., \cite{saguy2008fat, ravary2019shaping, greenberg2003portrayals, herbozo2004beauty, ata2010weight}). These meanings are also now pervasive in private culture, as documented implicit association tests, explicit attitudes, vignette studies, and even in neural responses to images of people with varying body shape (e.g., \cite{nosek2007pervasiveness, ravary2019shaping, carels2013stereotypical, teachman2003demonstrations, teachman2001implicit, hinman2015stereotypical, schwartz2006influence, schupp2011implicit,azevedo2014weighing, lieberman2012disgust}). We focus on four specific meanings attached to obesity: those around gender, morality, health, and social class. 


Fat is gendered in both its cultural connotations and social implications. For women, the aesthetically ideal weight is harder to achieve and more narrowly defined than it is for men \citep{fikkan2012fat}. Fat women are disproportionately stigmatized for their weight compared to fat men \citep{fikkan2012fat}. Because female bodies are more closely scrutinized and policed than male bodies \citep{fredrickson1997objectification, bar1976physical, jackson1992physical}, fat (and weight control more broadly) becomes implicitly associated more with women than with men. In fact, fat men are perceived as more effeminate than leaner men. This feminization is epitomized by gendered colloquialisms like "man breasts," "bitch tits" \citep{bell2007feminism}  and "moobs." By contrast, muscularity and leanness signal masculinity \citep{stearns2002fat}. 

Shared meanings around weight also include moral judgments. Amid surging American excess and consumerism between 1890 and 1910, lean body weight emerged as a visible marker of self-restraint \citep{stearns2002fat}. This moral dimension of fat schemata persists to the present day \citep{schwartz2006influence,saguy2005weighing,abigail2013s,saguy2008fat,saguy2010morality}. Fatness connotes gluttony, sloth, laziness, and a lack of self-control. Slenderness, by contrast, signals restraint and self-discipline. In turn, widespread belief that body weight is under individual control constructs fat as an external manifestation of an internal character flaw. The gravity of this judgment is exaggerated by the value placed on agency and free-will in the United States \citep{crandall2001attribution}. 

As excess body weight was constructed as a health condition in the 1950s \citep{sobal1995medicalization}, the aesthetic, moral, and medical meanings of fat mutually reinforced each other. Following the 1950s, body weight surged \citep{biltekoff2013eating,stearns2002fat}, and middle class women increasingly demanded that their physicians help them lose weight. Even if women's motivations were primarily aesthetic \citep{stearns2002fat} and driven by the gendered consequences of being overweight, doctors' roles expanded to encompass weight management. The moral undertones of excess body weight were reinforced by the medical profession's emphasis on behavioral causes and behavioral solutions, namely caloric restraint and increased expenditure through exercise \citep{stearns2002fat}. As fatness became medicalized and hence viewed as a physical illness, it picked up additional stigma. Illnesses (both physical and mental) are stigmatizing, and garner even more stigma when viewed as controllable \citep{crandall1995physical}, because self-control is constructed as a matter of character and morality. By emphasizing controllable factors and individual responsibility, the medicalization of fat in the 1950s likely contributed to its stigmatization. 

In the latter part of the 20th century, biomedical scientists found that heavy body weight was correlated with the risk of various diseases. These findings reinforced the perception that excess weight was a health condition. Then, in the early 1970s, the Body Mass Index (BMI) emerged as a standard (but crude) proxy measure for body fat \citep{blackburn2014commentary}. Continuous BMI values were mapped into discrete categories using cutoffs whose values are based on correlations with diseases \citep{kuczmarski2000criteria}. These BMI-based categories thereby characterize individuals' weight status alongside correlated health risks. Typical categories include "underweight," "normal" or "healthy," "overweight," "obese," and "morbidly obese" \citep{kuczmarski2000criteria}. Such category labels explicitly mark someone as normative or deviant. Just as labels like “criminal” or “mental illness” guide our understanding of deviant behavior \citep{ervinggoffman1963notes}, labels for deviant health and weight status frame how we think of our own and others' bodies. The moral imperative to achieve low body weight was thus reinforced by the emerging and quantifiable conviction that excess weight led to disease.\footnote{Despite correlations between BMI and disease, the causal relationship between weight and health remains weakly understood, and is a debate of its own \citep{campos2005epidemiology}. We are primarily concerned with the \textit{perception} and widespread cultural understanding that excess weight causes poor health.}

In the 1990s, America’s weight "problem" was reinterpreted as a public health crisis: a societal issue rather than merely a medical condition \citep{saguy2005weighing}. As more and more Americans were categorized as overweight or obese, the government declared war against fat with the 2001 publication of the Surgeon General's \textit{Call to Action to Prevent and Decrease Overweight and Obesity} \citep{office2001surgeon}. This document highlighted the heavy "burden" of overweight and obesity on the United States, in "premature death and disability, in health care costs, in lost productivity, and in social stigmatization" \citep{office2001surgeon}. Surgeon General Carmona restated this position more bluntly in 2003: "While extra value meals may save us some change at the counter, they're costing us billions of dollars in health care and lost productivity. Physical inactivity and super-sized meals are leading to a nation of oversized people" \citep{carmona2003obesity}. By labeling obesity as a public health crisis, government agencies constructed weight management as a civic duty, in addition to a moral imperative and a personal responsibility \citep{biltekoff2013eating,abigail2013s}. 

Our understanding of weight is also intertwined with socioeconomic status (SES). Historically, fatness signaled prosperity while thinness signaled starvation, poverty and illness \citep{ernsberger2009does}. In contemporary America and other affluent societies, however, this association is reversed \citep{ernsberger2009does}. Thinness is now a source of status distinction. It is particularly prized among the affluent, and especially among high-SES women \citep{ernsberger2009does}. Thinness signifies access to healthy meals and to exercise; such access costs time and money. Fatness, by contrast, is construed as a problem of poor people, racial and ethnic minorities, and immigrants \citep{herndon2005collateral}. Because low social class already connotes laziness under prevailing stereotypes, moralistic understandings of obesity as a "poor person's problem" are mutually reinforcing. Thus, public health campaigns against fat Americans implicitly blame low SES individuals, and, in turn, racial and ethnic minorities. Indeed, some scholars describe the "war against obesity" as a war against low-class and minority individuals \citep{campos2005epidemiology,herndon2005collateral}.

Public culture is a key site for both the construction and transmission of meanings around body weight (e.g., \cite{grabe2008role, lopez2010influence, ravary2019shaping, lin2009relationship, ata2010weight, latner2007childhood}). In particular, news reporting presents a critical site for meaning-making around obesity, as it conveys scientific and governmental understandings of obesity to a mass audience \citep{boero2007all,saguy2008fat, saguy2014reporting}. Between 1980 and 2004, news coverage of obesity increased exponentially, following a similar surge in scientific publishing on obesity \citep{saguy2005weighing}. An article might describe government efforts to combat obesity or epidemiological studies on the prevalence – and correlates – of obesity. In the process of reporting, information is framed in ways that suggest specific interpretations and transform the original meaning. For example, in 1994 headlines focused on a study published in the \textit{Journal of American Health} that documented the rising prevalence of overweight and obesity. Press coverage of that study popularized the metaphor "obesity epidemic" \citep{saguy2005weighing} — a metaphor that is now taken literally \citep{mitchell2007us}. In an experimental study, \cite{saguy2014reporting} found compelling evidence that way that news articles frame obesity can significantly increased individuals' anti-fat prejudice, compared to a control article. As these examples illustrate, news reporting on fatness reflects and contributes to patterns of large-scale private meaning-making around body weight. However, it remains unclear exactly how public constructions of body weight become encoded by individuals. In fact, this is an important case for a more generic theoretical puzzle: how do we learn private culture from public culture?

\vspace{5mm}
\begin{center}

\item{\section*{CULTURE, COGNITION, AND COMPUTATIONAL MODELS}}

\end{center}

\section*{Cognitive socialization through schemata and frames} 

Previous literature has focused on the framing of obesity: how cultural objects like news reporting guide the interpretation of obesity as gendered, immoral, unhealthy, and low-class. We aim to close the theoretical circle implicit in this scholarship, by building computational models of the schemata learned from and invoked by the frames in news reporting. More generally, we want to understand how public culture becomes private culture. To do so, we build on the literature on frames and schemata. First, we tease apart these two often conflated terms and clarify the part each plays in cognitive socialization and everyday meaning-making \citep{wood2018schemas}. We then arrive at a plausible computational strategy for modeling schemata. 

In previous work, a frame is described as a "structure" \citep{goffman1974frame} that filters how a cultural object is interpreted, such as the collection of euphemisms, stereotypes, and anecdotes in a news article that guide its interpretation. The term "frame" is typically applied to structures of meaning that exist external to individuals, such as in news articles. Frames parallel the concept of "schemata," which are mental structures that organize individuals' mental life \citep{dimaggio1997culture}, and the  "building blocks" of cognition and knowledge \citep{rumelhart1980schemata}. Broadly, schema theory provides an account of how knowledge is represented and how these representations are manipulated. Thus, while frames describe ensembles of meaning external to the individual, schemata describe internalized ensembles of meaning \citep{lizardo2017improving}. 

Both schemata and frames refer to structures and processes that allow us to generate and interpret our cultural environment. To be clear, making meaning from a cultural object requires an interaction between frames and schemata: we are guided externally - by the features of the object -  and internally, by our preexisting schemata. In the special case where there are no preexisting schemata to guide learning (i.e., learning from a blank slate), the schemata learned from a cultural object may closely match the frames in that object. In other cases, features of the object activate preexisting schemata, while our preexisting schemata simultaneously tune us to certain features of the object.

Both schemata and frames may also be hierarchically structured: a higher level frame may be an ensemble of lower-level frames, and schemata may comprise lower level schemata. For example, a higher-order schemata around the process of "going on a diet" might be comprised of lower-order schemata such as  "fat is unattractive," and "fat is unhealthy," which in turn are comprised of relatively lower-order schemata around "fatness," "attractiveness," and "health." In other words, the concept of going on a diet is scaffolded from these relatively lower level concepts and relations between them. In this paper, we aim to get at schemata for words and schemata for four dichotomous concepts: gender, morality, health, and social class.

Frames and schemata are sometimes used interchangeably or imprecisely in the sociological literature. In this paper, we strictly differentiate between frames and schemata. Schemata refer to any internal cognitive structure used in meaning-making. Frames, by contrast, refer solely to external aspects of the cultural environment. For example, when referring to the interpretation of a frame, we are really talking about schematic processes activated by the frame; interpretation is a cognitive process that unfolds through the interaction of internal and external resources. While there is consensus that both frames and schemata are core to organizing the social world, there remains a lack of consensus on the nature of their "structure" and processes. As we will see in the next section, to bridge this gap in cognitive sociology we need to provide an account of schemata that is a more plausible at cognitive and neurological levels of detail.\footnote{Here we differ from \citet{wood2018schemas}. They view the vast majority of schemata as cultural. We follow \citep{foster2018culture} in reserving the modifier cultural for schemata that are widely shared. We also require that their similar content arise from participation in social life; more explicitly, through social (i.e., inter-personal) cognitive causal chains. Many low-level schemata (e.g., those having to do with perception or motor activity) may be widely shared, but they are shared due to common experience of the natural world or because they are part of ``core knowledge" built into the human brain by evolution.} 

\begin{raggedright}
\section*{Operationalizing Cognitive Sociology in Computational Models}
\end{raggedright}
How do we develop a useful and plausible (even if not completely realistic) account of schematic processing? To frame our answer to this question, we draw on computational neuroscientist David Marr’s three "levels of explanation" for information processing systems \citep{marr1982vision}. Moving from the most abstract to the most concrete, these are the computational, representational/algorithmic, and hardware levels. These levels move from abstract to concrete \citep{foster2018culture,marr1982vision}. While the levels of analysis are conceptually distinct, they are interrelated; often, a model aims to capture aspects of a phenomenon at multiple Marr levels (as we do here).

The \textit{computational level} focuses on what problems are being solved by the information-processing system,  why they are being solved, and the broad strategy for solving them. To take Marr's classic example, given a set of prices, a cash register solves the problem of determining what the customer should pay by outputting a bill \citep{marr1982vision}. A cash register's strategy is to sum up the prices; it does this rather than multiplying the prices or combining them in some other way because the rules we consider fair for combining prices actually define the mathematical operation of addition. On the conventional sociological understanding, topic modeling extracts frames from text, assuming a generative model for documents in which writers first select combinations of frames to weave together, and then from each frame use some combination of language corresponding to a frame generate a document \citep{dimaggio2013exploiting}. The computational problem being solved in the "forward" or generative direction is to produce a document from a set of frames; in the inverse, inferential direction, topic modeling retrieves the frames used to generate documents \citep{jones2015models}.\footnote{Recall that frames are "groups of words that are associated under a common theme" or "semantic contexts that prime particular associations or interpretations of a phenomenon in a reader" \citep{dimaggio2013exploiting}. In our distinction between frames and schemata, topic modeling might be better understood as recovering the \textit{schemata} that generate text, as well as the schemata invoked by text, rather than frames. This is because topic modeling captures the cognitive generation, and reception, of text data; the frames would then correspond to the aspects of the text that lead to the activation of particular schemata (i.e., "semantic contexts that prime particular associations or interpretations" \citep{dimaggio2013exploiting}). In our case, we are capturing schemata because we are modeling the process of constructing semantic memory from experience (in this case, learning the meaning of words from text).} 

In this paper, the computational problem is how public culture becomes private culture, and the broad-strokes strategy is the learning of schemata. More specifically, we address the problem of learning schemata related to word meanings by reading text data. These schemata are a form of \textit{semantic memory}: "memory for word meanings, facts, concepts, and general world knowledge" \citep{jones2015models}. Since word meanings gain their content through participation in social life, the building blocks of this kind of semantic memory are cultural schemata. Here as elsewhere, humans rely on schemata because they compress incoming information and efficiently process new information. Cultural and cognitive sociology tend to hover at this Marr level: theorizing about the functions and presence of schemata. Thus, while sociologists broadly agree that schemata are key structures for organizing our knowledge, there is little discussion of what these structures entail. In Marr's language, sociological work on culture and cognition ruminates at the most abstract level.

Going more concrete, the \textit{representational/algorithmic} level models what representations are used in a system and how these representations are manipulated as the system moves from input to output. These questions are intertwined; how information is represented affects how it may be manipulated \citep{marr1982vision}. Both the cash register and the cashier use numerals to represent quantities and algebra to manipulate these representations; this provides an explanation of the cash register's behavior at the representational/algorithm level. In topic modeling, documents are represented as probability distributions over frames and frames are represented as probability distributions over words. In the generative direction, sampling from these distributions suffices to generate a document. In the inverse direction, a range of inferential algorithms (e.g., Latent Dirichlet Allocation) are used to recover these distributions from raw text. Topic modeling does not necessarily aim for these representations, sampling processes, or inferences to be cognitively plausible for a writer generating a text document or a reader uncovering the latent frames. The representational/algorithmic level is rarely addressed in sociology (at least, explicitly); it is usually the focus of cognitive psychology. In this paper, we contribute to cultural and cognitive sociology at this Marr level by modeling schemata related to word meaning with distributed representations, and modeling schema acquisition by training a neural network through backpropagation. As will be explained shortly, a distributed representation is the idea that a concept, such as "fat," may be represented by its unique pattern of connection to a limited number of low-level units (e.g., neurons). We also describe how these schemata may be learned and activated by external sources like news. 

The \textit{hardware level} is, for human beings, usually the domain of neuroscientists, since modeling focuses on the physical implementation of the information processing system. The cash register diverges from a cashier at this level; the parts of a cash register do not resemble the human brain, but they do not need to do so in order to fulfill the purpose of producing a valid bill. Similarly, a topic model does not aim to model how hardware in the brain tracks and manipulates probability distributions, only to extract topics and frames from text data. We aim to contribute to understanding culture at this Marr level by providing an account of representations and algorithmic manipulations that is more consistent with human hardware. 

Empirically, we investigate various schemata around obesity with word2vec. First, we extract distributed representations for keywords around obesity and health, such as "obese," and "slender," and explore their meanings. Second, we extract slightly higher-order schemata: distributed representations of the dimensions of gender, morality, health, and socioeconomic status. These are dimensions in that they are defined through an opposition – for example, masculinity exists in opposition to femininity, and immorality exist in opposition to morality. We then show how obesity and related concepts are refracted through these dimensions, recovering the cultural connotations discovered by qualitative scholars over the past two decades.

\section*{Connectionism and Distributed Representations}

To represent schemata, we draw on the connectionist tradition in cognitive science \citep{rumelhart1986sequential} and cultural sociology \citep{strauss1997cognitive}. Connectionists theorize mental representations and processes as networks of associations. This point of view suggests, for example, that when we see the word "obese," we do not immediately retrieve the concept obesity from a unique slot; rather, viewing these letters triggers a pattern of activation in a limited number of units (e.g., neurons, or groups of neurons) organized in networks. That pattern of activation, then, is a reconstruction of our concept for obesity, and is likely similar to patterns triggered by related words such as "overweight" and "slender" (which also refer to body weight and shape). Key to this approach is that 1) concepts' meanings are distributed across many units (e.g., neurons), an approach referred to as "distributed representations" and 2) individual units are involved in representing many concepts. \footnote{We have described connectionism with distributed representations. This is our focus in this paper; it is also the most common approach in contemporary cognitive science and machine learning. It is, however, possible to have localists representation in the connectionist framework; in that case, individual units (e.g., neurons) represent a specific concept and the network encodes associations between those concepts \citep{Thagard2012}.} 

Meaning is embedded not only in the patterns of activation but in the relationships between these patterns of activation: for example, whether the pattern for "obesity" is more similar to the pattern for "attractive" or "unattractive." Thus, for connectionists, schemata come in the form of distributed representations and relationships between them.  

Connectionism stands in contrast to the symbolic approach to represent meaning, which theorizes the mind as a "file cabinet" or "storage bin" where each concept is a distinct entity – or symbol –  stored in isolation from other concepts \citep{smith1996connectionism}. In comparison with the symbolic approach, a distributed representation is a more plausible strategy to store concepts at Marr's representational/algorithmic level of analysis. First, it is more memory-efficient. In contrast to maintaining slots to represent all possible concepts in the world, storing a concept as a distributed representation means that many concepts can be represented by a limited number of neurons, since even a limited number of neurons may have many, many possible patterns of activation \citep{goodfellow2016deep}.

Distributing a concept across many units also enables knowledge to be fuzzily defined, flexible, and robust. Even if one connection or unit is severed or not activated, a concept can still be reconstructed by the many remaining units. This also suggests how mental reconstruction of concepts, such as "thin," can vary according to contextual stimuli (e.g., frames) and still roughly refer to the same external phenomenon. For example, in popular women's magazines, the pattern of activation for the concept "thin" likely comprises a slightly different distributed representation than reading a scientific article on anorexia nervosa. At the least, "thin" might activate positive valence in the first and negative valence in the latter. However, both refer largely to the concept of "thin" as a low proportion of body fat. 

Connectionism provides a plausible account for learning. A schema can be learned, or altered, by using existing patterns of activation as a blueprint. For example, repeated exposure to a representation of “thin” in magazines where it is framed as beautiful might strengthen the co-activation between the concepts of "thin" and "beautiful." Then, even in situations where "thin" is not framed with respect to attraction, the concept of "beautiful" will be activated alongside "thin," as they now share similar patterns of activation. This is called spreading activation, or semantic contagion \citep{ross1992semantic, collins1975spreading}.

More generally, schematic processing might occur in two ways \citep{rumelhart1980schemata}. First, from the top down, where the activation of higher-order schema also (partially) activates other sub-constructs. For example, when we envision an "obese person" many aspects of the distributed representations for "waist," "body," etc. are also likely activated (or partially reconstructed), since our schemata for "obese persons" is comprised of these sub-concepts. In other words, given a big picture, we fill in information.  Some predictions, of course, may be erroneous, such as whether this person is rich or poor. Filling in this bigger picture with details is a heuristic, enabling speedy processing at the risk of decreased accuracy. 

Second, schematic processing may occur bottom up, enabling us to complete partial stimuli to gain a larger picture. Here, we are also predicting missing information to gain a more complete picture. For example, consider information that a young individual is severely underweight, and is seeing a therapist and a dietitian. From such partial information, we may quickly fit this to a larger schema, such as anorexia nervosa. Upon arriving at this big picture, we might (possibly incorrectly) then use top-down processing to fill in missing details to our schema for anorexia nervosa, such as that this person is female (rather than male) and wealthy (rather than poor). Research in race and gender investigate such bottom-up and top-down schematic processing as implicit associations and stereotypes.  

The connectionist cognitive architecture is supported by empirical studies as well \cite{Thagard2012}. For example, viewing various semantic categories of images and words accurately predicts distinct spatial patterns of neural activation \citep{mitchell2008predicting,mcclelland2003parallel}. Further, research asking individuals to list important features of concepts yields a core set of features listed consistently across individuals \citep{cree2003analyzing}. 

While connectionism is often contrasted with symbolic approaches, these two frameworks are in fact complementary. For example, language is symbolic in that our words are discrete representations, but meaning is continuous. This means that understanding language requires us to translate words (symbolic) to meaning (distributed). 

As we will show, neural word embeddings also represent and manipulate knowledge according to symbolic and connectionist strategies. Thus, embeddings are cognitively plausible models of schematic processing at the representational/algorithmic Marr level of understanding. For plausible hardware implementation of schematic processing, we turn to artificial neural networks as very simple analogies to biological neural networks. As we explain, in artificial neural networks, a concept (distributed representation) is defined by the weighted connections to artificial neurons \citep{smith1996connectionism,goodfellow2016deep}.

\vspace{5mm}
\begin{center}
\item{\section*{MODELING MEANING WITH ARTIFICIAL NEURAL NETWORKS}}
\end{center}

Research in distributed representations, when combined with an approach to machine-learning called artificial neural networks (ANNs), has yielded a strategy to model the meaning encoded in language, called word2vec. Word2vec also stands out from other computational methods to analyze text for its performance on linguistic tasks, such as recognizing analogies and synonyms, and because it mirrors aspects of human cognition \citep{gunther2019vector}. 

\section*{Artificial Neural Networks}

An ANN is a computational model used to detect patterns in data for some task like classification. An ANN makes predictions from data through a series of linear and non-linear transformations of the input. In contrast, regression methods make predictions from a single, linear or non-linear, transformation of the input. 

ANNs are inspired by biological models of neural networks. The basic building block of ANNs is an artificial neuron. Just as a biological neuron receives data from other neurons and has a synapse which "fires" according to some activation function, an artificial neuron receives, transforms and outputs data. \href{figure:1}{Figure 1} visualizes an artificial neuron in red to show how it receives weighted input data.

\begin{figure}
\label{figure:1}
\caption{Artificial Neural Network with a closeup of a neuron in the hidden layer}
\begin{center}
\includegraphics[scale=.5]{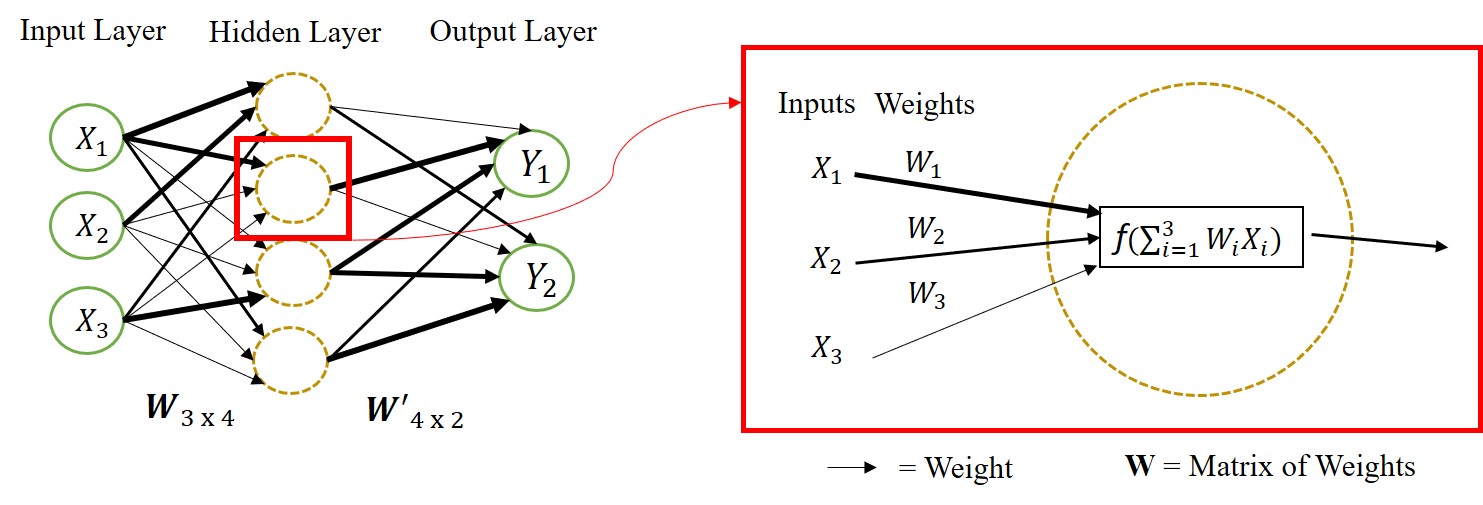}
\end{center}
\end{figure}

In fact, logistic regression is a simple ANN with just one neuron and a logistic activation function, which makes one prediction. In logistic regression, input data is weighted (with "coefficients") and then this weighted sum is passed to a logistic function – which "fires" at a given threshold to yield 0 or 1. 

While logistic regression only includes one artificial neuron and one predicted outcome, ANNs may include any number of neurons and predictions. Each neuron can weight incoming data differently and activate at different thresholds according to different functions. Furthermore, in "deep" neural networks, the outputs from the one set of neurons (a "layer" of neurons) then feed into the next set of neurons. While each neuron performs a simple task, it is easy to imagine how ensembles of neurons can perform complex tasks - like "ants building an anthill" \citep{geron2017hands}.

As another simple ANN model, consider an ANN consisting of three layers across which data is inputted and transformed to make two predictions about a person, from 3 binary observed variables. (\href{figure:1}{Figure 1}). This can be understood conceptually or mathematically. Both stories begin in the input layer, where data is inputted as a vector. For example, a vector describing three binary variables ($X_1, X_2,$ and $X_3$) about a person might be (0,0,1). This vector is called 3-dimensional since it contains 3 numbers. 

Conceptually, each of the four neurons in the second (hidden) layer then weights the three variables to yield a weighted sum of the observed variables. Then, each these four neuron "fires" according to an activation function (such as a logistic function) applied to that weighted sum. In the case where we apply a logistic function to each weighted sum, each neuron spits out a number between 0 and 1. With four neurons in the hidden layer, this yields four numbers. This list of numbers is now interpreted by the third (output) layer which consists of two neurons and no activation function. Each of these two neurons again weights the incoming four numbers from the hidden layer, and spits out a weighted sum for $Y_1$ and a weighted sum for $Y_2$ as the ANN's predictions. 

This can also be understood mathematically, rather than conceptually. First, the inputted vector is multiplied by a matrix of weights, yielding the element of a new vector. This vector is now passed to the second layer, where each element is transformed with some non-linear function, such as a logistic function. Now, this vector (which consists of weighted, summed, and then transformed data) is again multiplied by another weight matrix, and rows are summed, to produce predicted values or categories in the output layer. Note that only the second layer transforms the data non-linearly, while the input and output layers just pass through information inputted into them. The length of the vector passed to the hidden layer is set by the researcher, but is usually less than the dimensionality of the inputted data itself.  

\href{figure:1}{Figure 1} illustrates a neural network architecture with a single hidden layer, where the hidden layer has 4 dimensions, and predicts two outputs. ANNs may include more than one hidden layer, but for the purposes of this paper we focus on ANNs with a single hidden layer, sometimes referred to as shallow (versus deep) learning. Note that not all ANNs apply non-linear transformation(s) at the hidden layer. Indeed, word2vec does not apply a non-linearity; each neuron in the hidden layer simply passes through the sum of weighted inputs.

An ANN is trained on data to find values of weights that yield the highest performance according to some objective function. This can be thought of as learning to weigh the evidence for a prediction outcome. Optimal weights are learned though the iterations of two steps. First, data is inputted ("fed forward") into the ANN to predict an output, and the prediction error is calculated. Second,  each of the weights in the ANN are updated to minimize this prediction error. The first time these steps are run, the weights may be randomized.\footnote{As with many practical elements of deploying ANNs, there is an art to choosing such \textit{initializations} Randomizing weights is a simple but not necessarily optimal choice.}. Among the various algorithms to find a set of weights that minimize the prediction error, the most commonly used algorithm is backpropagation with stochastic gradient descent. Since an ANN is not a system of linear equations, traditional calculus methods (i.e., setting a derivative of the error function to 0 and finding the minimum) cannot be used to determine the combination of weights that minimizes the error. 

Backpropagation with stochastic gradient descent offers a solution. Specifically, at a given combination of weights, we calculate how the error increases or decreases if we slightly increase or decrease each weight. In more technical terms, we are finding the derivative (slope, or gradient) of the error function with respect to each weight. We can then update each of the weights, layer by layer starting from the layer closest to the predicted output, by changing the each weights by a small amount in the direction that reduces their error. The amount that we update the weights is called a ``learning rate'' and is set by the researcher. To deploy a common analogy: Much like trying to find the lowest elevation while walking in a terrain full of valleys, it can be tricky to know if a set of weights is a local minimum (which greatly reduces error) or global minimum (which yields the lowest possible error). To avoid getting stuck in a local minimum (e.g., a ditch rather than lowest valley), some small amount of randomness (stochasticity) accompanies the direction that the weights are updated.

ANNs were originally inspired by models of the brain. Literature on neural networks and neuroscience have diverged in recent years but many parallels persist, as do many opportunities for these literatures to again converge \citep{marblestone2016toward}. For example, just as an ANN is trained to optimize error functions by adjusting weights, it is likely that the brain optimizes error functions by adjusting processes and structures such as synapse firing \citep{marblestone2016toward}. The parallels between ANNs and our own cognition offer a response to DiMaggio’s call to integrate culture and cognition \citep{dimaggio1997culture} through application of ANNs to cultural data, such as news discourse about obesity. This integration goes beyond DiMaggio's call by attending more closely to all three Marr levels, and drawing on models that are plausible at the computational, representational/algorithmic, and hardware levels. 

\section*{Word2vec}

Word2vec uses an ANN to learn distributed representations of words from a text dataset. These distributed representations are the main goal of using Word2vec. Specifically, each word is represented as a vector (word-vector) with each number in the vector corresponding to weights to the hidden layer.\footnote{Equivalently, to the activation of the hidden layer triggered by a particular words, \citep{gunther2019vector}.} Alternately, we can think of word-vectors graphically - as vectors in $N$-dimensional semantic space (the representation of a collection of words as a collection of points in a semantic space is also referred to as a word embedding). For example, in a two-dimensional semantic space (a plane) a word-vector would have two elements to specify its position on the $x$-axis and $y$-axis. 

In word2vec, word-vectors are $N$-dimensional, where the dimension of the embedding space $N$ is set by the researcher. If words have similar meaning, they lie closer in semantic space, and have more similar word-vectors.\footnote{In a word2vec model, the $N$ specific basis vectors ("axes") picked out by the learning process are arbitrary, but the word embeddings encode a (latent) coordinate system in which dimensions correspond to salient dimensions of language. Meaningful directions in the semantic space, such as direction representing gender (masculine to feminine), are likely to correspond to some linear combination of the basis vectors (dimensions) initially constructed by the learning process.} The similarity between two vectors can be measured by cosine similarity. See appendix~\ref{appA} for a quick review of vectors and cosine similarity. 

Word-vectors found by word2vec are distributed representations in that each word is defined by where it lies on $N$ dimensions. The number of dimensions is usually between 100—1000, and thus is far less than the vocabulary size of the dataset itself. This makes the embedding a more efficient strategy to store words than "slots" for all possible vocabulary words. Each element in a word-vector also corresponds to a weight in a hidden layer in a neural network. Thus, each word-vector may also be loosely thought of as representing the how the word's meaning is distributed across lower-level, latent dimensions that organize language; or, alternatively, by how strongly the word activates different neurons in the hidden layer. In representing words as word-vectors, we are providing an account of (cultural) word-learning and semantic memory at the representational level - and our representational strategy is broadly consistent with the actual hardware implementation by which semantic knowledge is stored in the human brain.\footnote{As we shall see, the learning algorithm is also cognitively plausible \citep{gunther2019vector}.} 

To learn these word-vectors, word2vec gives a semantic prediction task to an ANN. While trying to successfully complete the prediction task, the ANN begins with randomized internal representations of words, and iteratively updates these representations. The first possible task is Continuous Bag of Words (CBOW), asking an ANN to predict a focal word given a set of context words. For example, the ANN is given the data excerpt "obesity is ? for health" and then predicts the missing word that was actually used in the data. 

In the example above, "obesity is ? for health," we use a context window of two; there are two words before the focal word (obesity, is) and two words after the focal word (for, health). This task is iteratively performed with various phrases in the data; as will be described, across many, many examples the ANN learns from mistakes to refine its predictive ability. The size of the phrases is specified by the researcher, and is called the "context window" and hereafter is also referred to as $C$. 

A second possible task, Skip-Gram, reverses this framework. Now, context words are predicted given a focal word. For the purposes of this paper, CBOW will be explained in more depth, although all concepts are largely similar across CBOW and Skip-Gram. Since word2vec with CBOW learns word-vectors by predicting words given their contexts, words that occur in similar contexts are represented with similar word-vectors.\footnote{It is important to note that word2vec is not merely learning word co-occurence. Word2vec is essentially an online algorithm for taking a word-context co-occurence matrix--weighted by the point-wise mutual information (PMI)--and finding low dimensional representations of words by matrix factorization \citep{levy2014dependency,gunther2019vector} In essence, the probability of word $a$ occurring in context $b$ is divided by the product their separate probabilities. This means that word2vec is sensitive not to \textit{contiguity}, i.e., raw co-occurence, but rather to \textit{contingency}, i.e., the degree to which $b$ is predictive of $a$ \citep{gunther2019vector}. This turns out to make word2vec more realistic at an algorithmic and hardware level \citep{gunther2019vector}} Word2vec is theoretically motivated by the linguistic Distributional Hypothesis, that words appearing in similar contexts tend to share meaning \citep{firth1957synopsis, harris1954distributional,gunther2019vector}. This hypothesis underlies many models of computational text analysis, including topic modeling, and is sometimes summarized as "a word is characterized by the company it keeps" \citep{firth1957synopsis}.

The neural network architecture of word2vec again includes a single hidden layer and the CBOW variant will be explained in detail (Figure 2). The story starts with small data excerpt: a set of $C$ context words from the data, such as "obesity is ? for health," where the focal word is known from the actual data. Conceptually, word-vectors for these context words (which are initially random) are first "activated" in the input layer. This means that among all possible word-vectors, only the ones that correspond to the context words in this data excerpt are passed into the hidden layer.\footnote{Mathematically, each context word is represented in the input layer as a vector with $V$ elements, one for each word in the vocabulary $
\mathcal{V}$, so $|\mathcal{V}| = V$. Each vector is "one-hot" encoded, meaning it is a vector of $0$'s and a single $1$; the element corresponding to the context word is indicated by a $1$, and the elements corresponding to all other vocabulary words are $0$. Thus, a context word-vector multiplied by the first weight matrix simply "selects" the vector of weights corresponding to that context word. Note that this one-hot encoding is a way to represent language symbolically, where each word has a "slot" represented by "1".} In the hidden layer, these context vectors are averaged, yielding a smoothed vector representation of the context. Next, the network computes the similarities (cosine similarity) between this context vector and another distributed representation for each vocabulary word, yielding $V$ similarities. This is a measure of how similar the context is with every possible vocabulary word. These similarities are then passed to the output via softmax (multinomial logit) yielding $V$ probabilities, each of which is the predicted probability that the $i^{th}$ vocabulary word is the focal word. Specifically, given a set of context words, the probability of seeing some output word $w$ is:

\begin{equation}
P(w,context)= \frac{\exp(score(w, context ))}{\sum_{w'=1}^{V} \exp(score(w', context))}
\end{equation}

where the $score(w_i, context)$ is the dot product of $i^{th}$ word-vector in the vocabulary and the averaged distributed representation for the context words. The denominator is the sum of exponentiated scores between context words and each word in the vocabulary.  

Note that there are two distributed representations manipulated by word2vec for each word. Each vocabulary word has a corresponding word-vector (which we later use) and a \textit{second} distributed representation (set of weights) from the hidden layer to the output layer, which (in the CBOW task) computes the score by taking the dot product of the context and each potential target word \citep{rong2014word2vec}. Word2vec adjusts the context word-vectors and the second distributed representation to maximize the probability of seeing the actual word missing from the data. In other words, word-vectors and the second distributed representations (i.e., the ANN weight) are tweaked to maximize the similarity between the context vector and the distributed representation for the actual focal world in the data excerpt, and minimize the similarity with other words.

The dimension of the distributed representation for each word-vector is the number of neurons. In Figure 2, the first set of distributed representations (word-vectors) is referred to as $W$, and has dimensions $V \times N$, i.e., vocabulary size times dimension of the embedding space. The second set of distributed representations is $W'$, and has dimensions $N \times V$. The output layer is another vector of $V$ dimensions.

\begin{figure}
\label{figure:2}
\caption{Artificial Neural Network for word2vec with CBOW}
\begin{center}
\includegraphics[scale=.5]{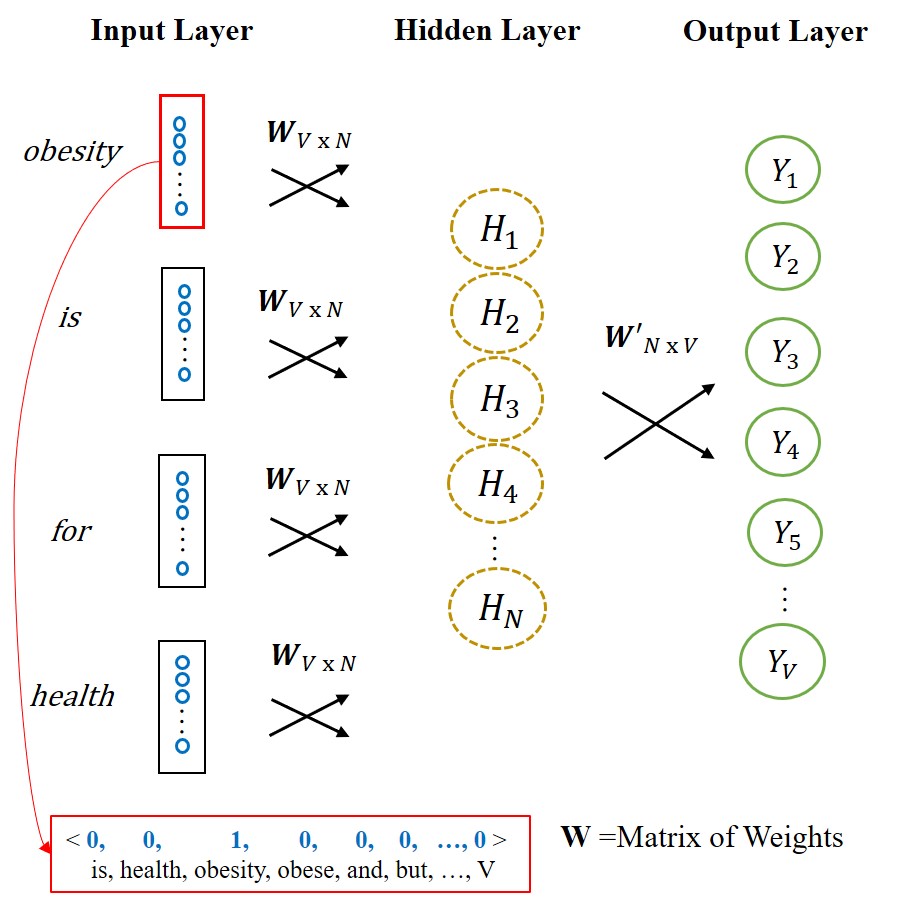}
\end{center}
\end{figure}

In practice, word2vec must approximate the softmax function because it is computationally intensive to directly compute the denominator at each pass. The two common approximations include hierarchical softmax and negative sampling. Negative sampling essentially trains the network to distinguish "positive" instances (i.e., the word used correctly) from "negative" examples (i.e., the same context with random words inserted) \citep{levy2014dependency,rong2014word2vec}. This latter strategy is a special case of a much more general approach called \textit{contrastive unsupervised learning}, which may have broad sociological application and interpretation \citep{arora2019theoretical}. For a more detailed account of the implementation of word2vec, see \cite{rong2014word2vec}.

Because we know what the the expected focal word is, given a set of context words in the data, we can calculate the prediction error, and proceed to adjust the weights (distributed representations) to minimize this error, following the same general approach described above for generic ANNs. This training can be done many times with each example of a context and target word in the data, to find a set of weights that minimizes prediction error and thus generates high-quality word-vectors. Once a word2vec model has been trained for sufficient predictive accuracy, we can extract the word-vectors for subsequent use. Although the word-vector for a particular word $w$ is derived from the weights that run from $w$'s representation in the input layer to the hidden layer (or, equivalently, the activation of the nodes in the hidden layer stimulated by that particular word), in practice we abstract word-vectors from their neural network origins and treat them as vectors in an $N$-dimensional vector space. Despite the simplicity of word2vec, as we show next, it can abstract from text data to learn rich and nuanced semantic meanings that match human intuitions.

\section*{Neural Word Embeddings, Culture, and Cognition}

We have described how neural word embedding methods (specifically, word2vec) learn semantic representations by "reading" text and making predictions. Now, we show that neural word embeddings are a cognitively plausible model for learning private culture (schemata) from public culture (news articles). To do so, we revisit Marr's three levels of analysis at which an information processing system may be understood \citep{marr1982vision}. We evaluate the cognitive plausibility of neural word embeddings as systems for schematic processing, at each Marr level. We focus on the word2vec model, but our arguments may be generalized to neural embedding methods more broadly.

We begin with the computational level: what problem is being solved, why, and with what strategy. In this paper, the computational problem we model is how public culture becomes private culture (schemata). More specifically, we describe the process of learning new schemata and reinforcing existing ones from "online" reading of text data. Just like human learners, neural word embedding are exposed to public discourse at large scale, and learns distributed representations of semantic information from this data (e.g., low-dimensional word-vectors). Both humans and neural embeddings learn schemata from the statistical experiences of stimuli (i.e., words) in their cultural environment in order to efficiently represent these experiences and efficiently process new experiences.

Word2vec was designed as a practical tool to work with natural language data at scale; it was not built as an explicit model of human cognition. But it is no coincidence that word2vec solves the computational problem of learning abstract and dense representations of word-meaning from language experience (which we humans also solve). Language modeling in computer science emerges from a lineage of efforts in psychology and cognitive science to model how humans process and represent semantic information (i.e., semantic memory) from distributional patterns in natural language \citep{gunther2019vector, mandera2017explaining}. One of the most well-known approaches in this lineage is Latent Semantic Analysis \citep{landauer1998introduction, landauer2007lsa}. Such models, often called distributional models, are not simply tools to process language: they were built to \textit{model} the computational problem of learning meaning from text data.\footnote{Describing a model as distributional refers to the distributional patterns of words and the use of the Distributional Hypothesis mentioned earlier, not to distributed representations. Word2vec is distributional but also uses distributed representations.}

A growing body of evidence demonstrates that word2vec, and other distributional models like Latent Semantic Analysis are compelling computational models of human meaning and semantic memory. For example, studies test how word-vectors perform on linguistic tasks that require deep semantic understanding, such as analogy completion and evaluating similarities between pairs of words \citep{mandera2017explaining, mikolov2013efficient, pennington2014glove, landauer1998introduction}. Word2vec stands out as cognitively plausible at the computational level because it reaches impressive performance on these linguistic tasks \citep{mandera2017explaining, mikolov2013efficient}. 

Numerous recent studies have also illustrated that the substantive meanings learned by word-vectors using these embedding methods corresponds powerfully to private meaning based on implicit association test, surveys, and even neural activity (e.g., \cite{caliskan2017semantics, bolukbasi2016quantifying, kozlowski2019geometry, grand2018semantic, jones2019stereotypical, joseph2020word, garg2018word, dehghani2017decoding, vodrahalli2018mapping}). For example, the cosine similarity between "male" and "science," vs between "female" and "science," correspond to the prevalent implicit associations of science as a masculine domain \citep{caliskan2017semantics}. Further, an abundance of work suggests that word embeddings correspond to neural information that is subconsciously activated when reading and hearing language (e.g.,  \cite{dehghani2017decoding, vodrahalli2018mapping}).\footnote{For example, \cite{dehghani2017decoding} trained embedding representations of English language narratives, using a generalization of word2vec (doc2vec, which learns distributed embeddings of documents as well as words). Using these embeddings, they were able to decode participants' neural activity (using fMRI scanning) corresponding to the reading of particular stories. Further, they could do so even when the narratives were translated to other languages and then provided to participants fluent in these languages.} 

Thus, empirical studies establish that the \textit{substantive} meanings that word2vec (and distributional models more broadly) learns correlate to private meanings. We build on this: we argue that neural word embeddings, including word2vec, additionally \textit{learn} these meanings in a way that is cognitively plausible, since they model human experience with public culture more closely than other distributional models \citep{gunther2019vector, mandera2017explaining}. The input to many distributional models (e.g., standard topic modelling) is a matrix of document-term co-occurences, and this matrix is then compressed using a dimensionality reduction algorithm, like singular value decomposition.\footnote{Other modern popular distributional models, like GloVe \citep{pennington2014glove}, also begin with a co-occurence matrix and then reduce the dimensionality of this matrix.} This is not cognitively plausible input: it would require us to build (and store) a very large co-occurrence matrix, and would require us to build this matrix before reducing it to arrive at condensed, more efficient representations \citep{gunther2019vector}. Upon experiencing new co-occurences of words, it would require us to completely re-do this dimensionality reduction. In contrast, the input to word2vec is a vast number of experiences with small excerpts of text, where meaning is learned iteratively from each experience. In short, word2vec learns in a more online fashion, similar to humans.\footnote{The fact that models like word2Vec specifically abstract from instances of contexts to deep word representations also models the transition from episodic memory to semantic memory, as noted by \citep{gunther2019vector} among others.} Taken together, this evidence illustrates that word2vec is a good model for learning schemata related to word-meaning from text data at the computational level.

The next Marr level of analysis specifies the representations that are used to learn private culture from public culture, and the algorithms used to manipulate these representations. Word2vec also carries many consistencies with human cognition at this Marr level.\footnote{In contrast, count-based models like Latent Semantic Analysis and GloVe are not cognitively plausible at this level, nor are they intended to be. For additional discussion comparing various language models, see \citep{gunther2019vector} and \citep{mandera2017explaining}. Our arguments around the cognitive plausibility of word2vec generalize beyond word2vec to neural embedding models more generally, like FastText \citep{bojanowski2017enriching}. We focus on word2vec given its pervasive usage.} Both word2vec and (plausibly) humans use distributed representations to represent and reconstruct learned knowledge. As previously illustrated in our discussion of distributed representations, distributing a concept across many units also enables knowledge to be efficiently represented, fuzzily defined, flexible, and robust. Word2vec manipulates and learns distributed representations using cosine similarity as a similarity metric, and maps from similarity to probability using the softmax function (familiar to sociologists in multinomial logistic regression). Humans also routinely use some similarity metric and map from a sense of similarity to a prediction, although they do not use the same representations or algorithms as those used by word2vec. For example, while cosine similarity weights all features of a distributed representation equally, humans place more weight on certain aspects of a concept \citep{chen2017evaluating}. The difference is that with this weighting, the similarity of two objects is based on their most salient aspects, rather than based on all aspects. For example, while the words "man" and "woman" have many aspects of meaning in common, we consider them very different because we heavily weight meanings of gender and these two words differ on gender. 

Further, just as human schemata may be decomposed into lower order schemata or built up into higher order schemata \citep{rumelhart1980schemata}, the representations in word2vec are compositional, to an extent. This is most famously demonstrated by the fact that word-vectors encode semantic regularities such as analogical relationships between words \citep{mikolov2013efficient,mikolov2013distributed}. For example, a word2vec model trained on Google News shows that the vector $\overrightarrow{king}$ – $\overrightarrow{man}$+ $\overrightarrow{woman}$ (i.e., removing the masculine component of $\overrightarrow{king}$ by "subtracting off" $\overrightarrow{man}$ and replacing it with a "feminine" component by adding $\overrightarrow{woman}$), is most similar to $\overrightarrow{queen}$. This is possible because, as we will show in this paper, word2vec encodes latent representations of gender. Word2vec has learned a higher order representation for gender which is scaffolded on top of words denoting masculinity and femininity, like "he," and "she."

Word2vec's algorithm for learning meaning is also cognitively plausible in another way: it learns distributed representations through predictively coding stimuli in its cultural environment. The ANN algorithm in word2vec learns words though attempts to predict, or fill-in, missing target words given context words with CBOW (or, to predict potential context words given a target word for Skip-Gram).\footnote{In fact, the difference between the CBOW and Skip-Gram tasks may be thought of as learning word-level schemata from top-down activation and bottom-up activation, respectively. The CBOW task is top-down, since a set of words are given and word2vec must predict how likely it is that each vocabulary word "completes" this partial context. The Skip-Gram task is bottom-up since a word is given and word2vec must predict which other vocabulary words might in the same context.} Just as word2vec learns by prediction, humans constantly engage in "predictive coding" as we try filling in missing information in our environment from prior and contextual knowledge \citep{lupyan2015words,rumelhart1980schemata, delong2005probabilistic, clark2013whatever}. Predictive coding works as a heuristic, enabling us to process information more efficiently and, sometimes, more accurately. Furthermore, brains engaged in predictive coding also likely adjust processes and structures such as synapse firing in response to errors, to minimize error of future predictions, analogous to word2vec \citep{lupyan2015words, delong2005probabilistic, mandera2017explaining, rumelhart1986sequential}. In particular, word2vec uses backpropagation with stochastic gradient descent to adjust word-vectors and thereby minimize prediction errors. Backpropagation with stochastic gradient descent is in fact a general case of a classic learning model in psychology, called the Rescorla-Wagner model \citep{miller1995assessment} as observed by \cite{mandera2017explaining}. 

In making predictions, humans may be guided by existing schemata previously acquired in their cultural environment; schemata are information (heuristics, or short-cuts) in themselves. As an example of a simple word-level schema, in hearing the word "programmer" we might assume that a male is being discussed, rather than a female, if we are used to more examples of male programmers than female programmers. Similarly,  as word2vec learns schemata from a data set where "programmer" tends to be described with masculine adjectives and pronouns, across iterations of training word2vec will learn to predict masculine words as surrounding "programmer." Like human learners, word2vec can pick up on schemata that are pervasive but not consciously detected by the human observer. For example, recent research emphasizes instances where word2vec learns certain occupations as masculine and others as feminine, in ways that match our own stereotypes \citep{bolukbasi2016quantifying,caliskan2017semantics}. Thus, meanings that are found using word2vec may reflect regularities in meaning across news data; these regularities may then be considered "cultural" schemata learned from reading news data. In short, word2vec and other neural word embedding approaches point toward a response to Goffman's call to ``form an image of a group's framework of frameworks— its belief system its `cosmology'  '' \citep{goffman1974frame}. Of course, since word2vec picks up regularities in word usage, it is smoothing over contexts rather than picking up nuanced uses of words.

Unlike humans, who accumulate schemata across a lifetime, word2vec begins at a randomized state before learning from the given data. Thus, we can model meanings as they are contained within an isolated field or discourse. One approach to pick up the schemata activated by stimuli in specific contexts is to restrict training data. We restrict training data to news reporting on obesity to isolate schemata that are learned or activated by discussions of fat, health, and body shape. 

We have now illustrated that word2vec is a cognitively plausible model for learning private culture from text data at the computational level, and highlighted the many consistencies at the representational/algorithmic level. At the hardware level, the ANN architecture of word2vec also shares certain similarities with biological neural networks. Like ANNs, biological neural networks are thought to operate by minimizing some costs (i.e., errors from predictions) \citep{marblestone2016toward}. Backpropagation as a learning algorithm for artificial neural networks may have correspondences in biological neural networks \citep{marblestone2016toward}. Of course, the cognitive plausibility of word2vec should not be overstated, particularly at the hardware level. For example, the scale of ANNs is much smaller than of biological neural networks; while we use a hidden layer with 300 artificial neurons in our models, a human brain has around 86 billion biological neurons \citep{herculano2012remarkable}. Still, word2vec is a step towards a cognitively plausible model of schematic extraction and the interpretation of text data. Further, it may be used empirically to model the schemata learned from public culture, such as news reporting on obesity. 

\section*{Data}

To empirically model the schemata around obesity learned from news, we collected NYT articles which were published between January 1, 1980 and July 15, 2016 that included at least one of the words: overweight, obesity, obese, diet, fitness, fat, fatty, fattier, or fatter. This strategy yielded 103,581 NYT articles. Articles were collected from the academic database LexisNexis. We limited data collection to a single newspaper to minimize potential variation caused by multiple papers' differing audiences and political orientation. 

We chose the NYT for several reasons. First, numerous core studies of obesity in the news use the NYT, and so using this same data enables us to compare our conclusions to these previous studies \citep{boero2007all,lawrence2004framing,saguy2010morality}. Second, like these previous studies, we selected NYT because it is an important case; it is a leading national news source with wide readership in the American public \citep{center2012changing}. Nevertheless, we also replicated methods and findings on the Google News pretrained embeddings (which are trained using a corpus that aggregates across various news sources). This suggests that the schemata learned from the New York Times may also be learned from or reinforced by other news outlets.

As part of our data cleaning, we looked for two-word expressions (such as "New York" and "obesity epidemic") based on collocation counts, using the built-in phrase detector in the Python library Gensim \citep{rehurek_lrec}. This enables word2vec to learn word-vectors for these expressions \citep{mikolov2013distributed}. For additional details on data cleaning, see appendix~\ref{appB}.

\section*{Implementation of Word2vec}

We trained word2vec models on our data using Gensim \citep{rehurek_lrec}. Training a word2vec model requires several hyperparameter choices \citep{rong2014word2vec, rehurek_lrec}. We chose the combination of hyperparameters that maximized performance on the Google Analogy Test (\href{table:1}{Table 1}). This is a standard strategy to evaluate the quality of word embeddding models and tests how well the model completes analogies \citep{mikolov2013efficient}. A query for analogy (such as king:man as ?:woman) can be answered by selecting the word-vector with the maximum cosine similarity to the vector resulting from $\overrightarrow{king} -  \overrightarrow{man}+ \overrightarrow{woman}$. If the model returns "queen," the analogy is correctly completed.  More formally, the word-vector that word2vec returns for an analogy query is: $\underset{x \in \mathcal{V}}{\mathrm{argmax}} (cos(x,\overrightarrow{king} -  \overrightarrow{man} + \overrightarrow{woman})),$ where $x$ is a word-vector in the vocabulary ($\mathcal{V}$). Our final model used 500 dimensions for word-vectors, a context window size of 10, negative sampling with 5 samples, and a context-bag-of-words learning architecture, as described earlier. Its performance on the Google Analogy Test is comparable to the original models \citep{mikolov2013efficient}. 

Another common quality check for word embeddings is to compare human ratings of word similarities with cosine similarities computed using a trained embedding model. A standard data set of human-ratings to evaluate this is WordSim353 \citep{matias2001placing}, which includes 353 similarities rated by at least 13 individuals. The human-rated similarities from this data set correlate to similarities produced by our main model (model A in \href{table:1}{Table 1}), with a Spearman correlation of .61 (p<.001). This is comparable to published word2vec models \citep{pennington2014glove}. Our model's ability to perform these linguistic tasks (similarities and analogy tests) also corroborates that it provides a good computational model for human meaning. For further details on implementation see appendix~\ref{appC}. 

\begin{table}
\begin{center}
\caption{Accuracy of four candidate word2vec models on sections of the Google Analogy Test}
\label{table:1}
\begin{tabular}{ m{3cm} >{\centering\arraybackslash}m{2cm} >{\centering\arraybackslash}m{2cm} >{\centering\arraybackslash}m{2cm} >{\centering\arraybackslash}m{2cm} }
  & & & & \\
  & \textbf{Model A} & \textbf{Model B} & \textbf{Model C} & \textbf{Model D} \\
  \textbf{Family Section} & 90 & 81 & 71 & 84\\
  \textbf{Syntax Section} & 60 & 59 & 56 & 54 \\
  \textbf{All Sections} & 57 & 63 & 55 & 60 \\
 \hline
 \multicolumn{5}{m{13cm}}{\textit{Notes:} Accuracy is the percent of questions a model correctly completes in the Family, Syntax, and all sections of the Google Analogy Test, respectively. Models vary by hyperparameters; all learn 500-dimensional word-vectors and a context window size of 10. Model A uses CBOW and negative sampling, Model B uses Skip-Gram and negative sampling, Model C uses CBOW and hierarchical softmax, and Model D uses Skip-Gram and hierarchical softmax. Model A represents our final parameter selection. Model A did not perform better with lower dimensionality or with a smaller context window.} \\

\end{tabular}

\end{center}
\end{table}

\section*{Word-Level Schemata in Obesity Discourse}

Our first goal was to understand how word2vec learned meanings of keywords related to obesity (e.g., obese, obesity, overweight). In \href{table:2}{Table 2}, we list the ten words with highest cosine similarity to selected keywords based on results from a model trained on all data. Because we describe specific words here and the words they are most similar to, we are focusing on word-level schemata. If we consider these word-vectors as activations of lower level units, the similarity between two word-vectors may also be thought of as their co-activation. 
    
Word2vec appears to capture our own intuitive understanding of these keywords related to obesity. For example, the closest word to "obese" was "overweight" and vice versa. "Obese" and "overweight" had a cosine similarity of .80 (possible range -1 to 1). For both "obese" and "overweight," the next most similar word was "underweight," and both had cosine similarities of .65 to "underweight." Initially, it may be surprising to see underweight as similar to overweight or obese. However, these two concepts share a lot of meaning, thus they \textit{should} hold similar patterns of activation. In other words, two opposite words are opposites precisely because they share most of their meaning except for a single distinguishing component, like heavy versus light body weight.\footnote{This also highlights a difference between how humans and word2vec manipulate meanings. As discussed earlier, while cosine similarity equally weights all aspects of meaning of two words, humans might differently weight a single distinguished component, like heavy versus light body weight, so that we do not consider overweight and underweight to have similar meanings.}

The five words closest to "obesity" were: "childhood obesity," "chronic disease," "obesity epidemic," "cardiovascular disease," and "heart disease." This suggests that the model has learned that obesity 1) has to do with health, 2) is specifically associated with these diseases, and 3) is a public health concern and is commonly referred to as an "epidemic" \citep{abigail2013s}. We also see that "obesity epidemic" is learned as similar to words like "economic crisis" and "aging population," suggesting that word2vec has learned "obesity epidemic" as a crisis. This might be expected when learning is restricted to news, which sensationalizes obesity \citep{abigail2013s}.

The words closest to "fat" tended to be words about nutritional meaning of fat such as "carb" and "trans fats" rather than body fat. It might make sense that a reputable national news source would primarily use "fat" in the context of nutrition rather than as a body descriptor; "fat" is generally a pejorative body descriptor. After examining various instances where "fat" is used in the raw NYT articles, we confirmed that it is commonly referred to in a nutritional or clinical context. 

\begin{table}
\begin{center}
\caption{Top 10 most semantically similar words and phrases to selected obesity-related keywords}
\label{table:2}
\begin{tabular}{ m{5cm}  m{10cm} }
 \hline
  \textbf{obesity epidemic} & epidemic, obesity, childhood obesity, aging population, arms race, economic woes, over-consumption, imbalance, economic crisis, economic slowdown  \\ 
 & \\
 \textbf{obese} & overweight, underweight, normal weight, malnourished, hypertensive, sexually active, severely obese, undernourished, anorexic, sedentary  \\ 
 & \\
 \textbf{obesity} & childhood obesity, chronic disease, obesity epidemic, cardiovascular disease, heart disease, malnutrition, cigarette smoking, weight gain, diabetes, eating disorders  \\ 
 & \\
 \textbf{overweight} & obese, underweight, normal weight, being overweight, anorexic, slightly overweight, sedentary malnourished, hypertensive, undernourished  \\
 & \\
 \textbf{fat}  & fatty, flabby, excess fat, carb, fat content, butterfat, saturated fat, flab, carbs, trans fat  \\
 & \\
 \textbf{slim}  & slender, skinny, tanned, shapely, lithe, petite, svelte, chiseled, thin, blond  \\
 & \\
 
 \textbf{skinny}  & scrawny, chubby, bald, slender, tall, thin, stocky, hairy, balding, slim  \\
 & \\
 
 \textbf{physically fit}  & physically active, disciplined, sedentary, competent, open minded, considerate, personable, fatigued, active, conditioned  \\
 & \\
 
 \textbf{anorexic}  & bulimic, emaciated, eating disorder, unattractive, anorexia nervosa, tomboy, anorexia, addict, alcoholic, adolescent  \\
 & \\
 
 \textbf{bulimic}  & anorexic, suicidal, irritable, self induced, seriously ill, bulimia, sexually active, diabetic  \\
 & \\
 
 \textbf{healthy}  & healthier, healthy diet, healthful, physically active, regular exercise, balanced diet, healthy lifestyle, unhealthy, low cholesterol, nutritious  \\
  & \\
 \textbf{unhealthy}  & unhealthful, undesirable, harmful, unbalanced, overeating, sedentary lifestyle, healthful, excessive, unappealing, unsafe  \\
 \hline
  \multicolumn{2}{m{15cm}}{\textit{Notes:} Words and phrases are ordered most to least similar to each keyword on the left, based on a word2vec model trained on New York Times articles from 1980-2016. Similarity is measured with cosine similarity.} \\
  
\end{tabular}

\end{center}
\end{table}

\section*{Capturing dimensions of gender, morality, health, and socioeconomic status}

Our next goals were to capture two higher-level structures of schemata invoked by news data: 1) dimensions representing gender, morality, health, and SES, and 2) schemata around obesity and body weight with respect to these four dimensions. We represent all schemata, again, with distributed representations. We build on established methodology \citep{larsen2015autoencoding,bolukbasi2016quantifying,caliskan2017semantics,kozlowski2019geometry} to do so, which involves two steps. First, extract a direction in semantic space corresponding to each dimension (e.g., a direction corresponding to gender). Second, compute how this direction makes up the meaning of an outcome word (e.g., how "obese" is gendered). 

To extract a direction in semantic space corresponding to a specific dimension, such as gender ($\overrightarrow{g}$) we first choose a set of "anchor" words representing femininity and masculinity. Consider first the simplest case, in which we pick a single anchor word for each end of the dimension (e.g., \textit{woman} and \textit{man} for gender). We then subtract the word-vector representing masculinity from its feminine word-vector counterpart:  $\overrightarrow{woman}-\overrightarrow{man}$. Conceptually, if we assume that the meanings of woman and man are largely equivalent (both human, adults, nouns, singular, etc.) except for their opposite gender components, then subtraction (largely) cancels out all but the gender differences across each component. This gender component is our gender direction $\overrightarrow{g}$: $\overrightarrow{g}$ points toward femininity and $\overrightarrow{-g}$ points to masculinity.

The resulting vector can be construed as an axis ranging from negative (masculine) to positive (feminine) with gender-neutral (or, more precisely, not distinctively masculine or feminine) at 0. In practice, this method captures the true gender difference alongside other differences (which are essentially noise for our purposes). To extract a less noisy gender difference $\overrightarrow{g}$, we can take the average of a multiple words representing femininity, and average of a multiple words representing masculinity, and then subtract these two averages. We refer to this method as the Larsen method, following \citep{larsen2015autoencoding}.

Upon extracting a direction, we can compute how the direction relates to a word-level schemata, that is, the relationship between these two distributed representations. This relationship is encapsulated by cosine similarity. For example, the cosine similarity between a word-vector, such as $\overrightarrow{overweight}$, and the computed gender direction provides a scalar corresponding to where "overweight" lies on the gender direction. A scalar of -1 corresponds to hypermasculinity, 0 to lack of gendered meaning that is distinctly masculine or feminine, and 1 corresponds to hyperfemininity. Conceptually, this scalar tells us how gender meanings contribute the overall meaning of "overweight" - both in direction (masculine or feminine) and magnitude (amount of masculine or feminine meaning).\footnote{In this case, a projection is mathematically equivalent to cosine similarity, since everything is unit normed (length of 1).}

\subsection*{Selecting Anchor Words To Extract Each Dimension}

To select anchor words to extract our dimensions, we followed different approaches, depending on the availability of a published lexicon corresponding to the particular dimension. For gender, we build on lexicon used in previous work extracting gender from embeddings \citep{bolukbasi2016quantifying, ethayarajh2019understanding}. For morality, we used an established, validated lexicon from Moral Foundations Theory \citep{haidt2009above}. Specifically, we used the words in this lexicon that represent purity/impurity, which is one of the five dimensions of morality developed in this theory and most salient to the moralization of body weight. For health and SES, we brainstormed an initial list of anchor words and then checked for completeness by examining the words with highest cosine similarities to these words in a trained word2vec model, and using a thesaurus. \footnote{We did not use words that describe body weight or shape as anchor words for the health dimension or any other dimension.} As described in the next section, we checked how robust our methods are to anchor word selection. For details on the selection anchor words, see appendix~\ref{appD}. 

\subsection*{Validating the Robustness of Our Methods and Findings}

Following the above methodology, we extracted four dimensions (gender, morality, health, and SES) in our word2vec model. We checked the robustness of our findings in four ways. First, we evaluated the accuracy of the dimensions at classifying words known to lie at one end of the dimension or the other, including the anchor words. Second, we evaluated how sensitive our dimension construction was to the selection of anchor words. Third, we examine sensitivity of our findings to the sampling of the data used to train word2vec model. Fourth, we evaluate the sensitivity of our findings to the particular method used to extract the dimension. Here, we summarize the key validation findings, and provide details in appendix~\ref{appE}.

Our core validation step is to examine the accuracy of this method at classifying words we strongly expect to lie at one end of the dimension or the other. For each dimension, we examined accuracy on anchor words used to extract the dimension, and tested accuracy on a fresh set of words that had not been used to find the dimension. Following machine-learning literature, we also refer to anchor words as training words. Across the four dimensions, accuracy at classifying the training set ranged 92\%- 95\%.\footnote{See \href{table:3}{Table 3} for total number of training and testing words for each dimension. To ensure that comparable and equal numbers of words on either side of the scale were used to extract a dimension, some training words were used multiple times as anchor words. For example, we used the word-pairs "wealth-poverty" and "affluence-poverty" when extracting SES. Testing words were all unique.} Accuracy at classifying the testing set ranged 92\% to 98\%. The fact that the testing accuracies are comparable to training accuracies suggests that extracted dimensions are generalizable beyond the training words we used to extract them. 

In another validation step, we use two alternate methods to classify words with respect to gender, morality, health and SES. Using these alternate methods yielded similarly high accuracy scores and corroborated our empirical findings. These two methods include 1) a second geometric method where we extract a semantic direction (which we refer to as the Bolukbasi method, following \cite{bolukbasi2016quantifying}) and 2) training a linear classifier to distinguish between anchor words for a given dimension. Classification accuracies and training/testing word counts for all three methods on the training and testing sets are shown in \href{table:3}{Table 3}. 

\begin{table}[hbt!]

\caption{Performance of Three Methods to Classify Words by 
Gender, Morality, Health, and Socioeconomic Status (SES) }
\label{table:3}
\begin{tabular}{m{1.5cm} >{\centering\arraybackslash}m{2.3cm}  >{\centering\arraybackslash}m{2.3cm} >{\centering\arraybackslash}m{2.3cm} >{\centering\arraybackslash}m{2.3cm} }
    & & & &  \tabularnewline
   & & & &  \tabularnewline
	& \textbf{Larsen et. al. Method} & \textbf{Bolukbasi et. al. Method} & \textbf{SVM} & \textbf{Total Words}   \tabularnewline
    \noalign{\vskip 1.7mm}  
	& \multicolumn{4}{c}{\textit{Training (Anchor) Words}}  \tabularnewline \cline{2-5}
  Gender & 109 (92\%) & 104 (88\%) & 114 (97\%) & 118   \tabularnewline
  Morality & 60 (94\%) & N/A & 60 (94\%)  &  64 \tabularnewline
  Health & 121 (93\%) & 104 (80\%) & 126 (97\%) & 130   \tabularnewline
  SES & 95 (95\%) & 92 (92\%) & 99 (99\%) & 100  \tabularnewline
  \noalign{\vskip 1.7mm}  
  & \multicolumn{4}{c}{\textit{Testing Words}}  \tabularnewline \cline{2-5}
  
  Gender & 57 (95\%) & 57 (95\%) & 59 (98\%) & 60  \tabularnewline
  Morality & 29 (97\%) & N/A & 29 (97\%)  & 30   \tabularnewline
  Health & 55 (92\%) & 57 (95\%) & 54 (90\%) & 60   \tabularnewline
  SES & 59 (98\%) & 59 (98\%) & 58 (97\%) & 60  \tabularnewline
 \hline
 \multicolumn{5}{m{14cm}}{\textit{Notes:} SVM= Support Vector Machine. SES= Socioeconomic status. Accuracy is based on a 
 word2vec model trained on all data. \
 The Bolukbasi method requires paired words, which is not applicable to our morality lexicon.}

\end{tabular}
\end{table}

Another core validation step examined the robustness of these accuracy rates and empirical findings to the sampling of news articles. Specifically, we trained 25 word2vec models using the hyperparameters of Model A and on a 90\% random subset of articles. We find that many of our results are robust to sampling and testing/training performance is robust to news article sampling, as described in more detail in the results section. 

Our extensive validation is partly motivated by concerns that empirical results from word embedding models may not be stable \citep{antoniak2018evaluating}--though other scholarship finds that conclusions from embeddings are robust to word selection, methods to extract dimensions, and corpus selection \citep{joseph2020word}. We, too, find that our results are robust across all these four facets of uncertainty. As mentioned in our data section, we also replicated our methods and results using a publicly available word2vec model which was trained on Google News articles.\footnote{Google News agglomerates news from various sources. This trained word2vec model is freely available \href{https://code.google.com/archive/p/word2vec/}{from Google}.} This step is a further robustness check on our methods, and we found comparably high accuracy rates to using the model trained on our New York Times data (testing accuracy ranged 95\%-98\% across the four dimensions using the Google model). However, this step also enabled us to understand how generalizable or contextualized our findings are to the New York Times. Overall, our results were corroborated on the Google News model, suggesting that New York Times is not a unique site for learning these schemata around fat.\footnote{For additional methodological details see our appendices and our \href{https://github.com/arsena-k/Word2Vec-bias-extraction}{code}.} 

\section*{Outcomes Measured: Relationships between dimensions and word-level schemata}

Our final outcomes of interest are the meanings of 36 obesity-related keywords. Most importantly, these include terms about fat and thin body shapes/weights. Terms for fatness include: "obese," "obesity," "overweight," "obesity epidemic," "flab," "flabby," "being overweight," gained weight," "weight gain" and "morbidly obese." Terms about slenderness include: "slim," "lithe," "thin," "skinny," and "slender." We also include terms reflecting concepts in our literature review, including: behaviors related to weight management (e.g., "dieting," "overeating," "sedentary," and "exercise"), health, eating disorders (e.g., "anorexia," "bulimia"), and muscularity (e.g., "muscular," "burly," "lean," and "toned").\footnote{We only include terms present in every validation model at least 40 times. As described earlier, observe that "fat" largely refers to the macro-nutrient rather than body shape.} As described earlier, these relationships can be easily encapsulated by the cosine similarity of a word with a dimension (yielding a numeric quantity). 

We report a specific result, such as whether "obese" is feminine versus masculine, as robust if it occurs across every single validation model. We opt for this stringent requirement over measures of statistical significance, as statistical inference on word embedding data is still an open research area \citep{ethayarajh2019understanding}. From literature reviewed earlier, we expect that our model will learn to connote obesity with women, immorality, illness, and low class.

\vspace{5mm}
\begin{center}
\item{\section*{RESULTS}}
\end{center}

\section*{Gender and Obesity} 
Our first dimension captures the archetypical concept for gender, ranging from masculine to feminine. Note that this dimension does not aim to capture the full variety of gender identity; rather, it operationalizes a specific gender schema we expect to be learned from popular culture - the masculine vs feminine dichotomy \citep{kachel2016traditional}. As described earlier, to define the meaning of a word with respect to a dimension in a given model, we compute the cosine similarity between the word-vector and the dimension. For example, the cosine similarity between a word-vector, such as $\overrightarrow{overweight}$, and our gender dimension provides a scalar corresponding to where "overweight" lies on gender dimension – how strongly this word connotes masculinity or femininity. A scalar closer to -1 suggests that this word connotes more masculinity, while a scalar closer to 1 suggest that this word connotes more femininity. Conceptually, this scalar tells us how gender meanings contribute to the overall meaning of "overweight" - both in direction (masculine or feminine) and magnitude (amount of masculine or feminine meaning). On our model trained on all data, 109 (92\%) of 118 training words, and 57 (95\%) of 60 testing words were correctly classified as masculine or feminine.

Recall that we create 25 distinct word2vec models with the same hyperparameters, but trained on distinct random samples of 90\% of the corpus. Using more stringent requirements for accuracy (in which a word is considered correctly classified  by our methods if it is correctly classified in every single model) this dimension correctly categorized 99 (84\%) of the 118 training words and 51 (85\%) of the 60 testing words in every single model. We refer to this more stringent standard as \textit{robust classification}. For example, "lad" and "mr" were robustly classified as masculine, while "feminine," and "goddess" were robustly classified as feminine (\href{figure:3}{Figure 3}).

In Figures 3-10, the average cosine similarity of a word with respect to a given dimension (averaged across the 25 validation models) is plotted as a point, in the middle of the standard deviation and range of cosine similarities across the 25 models. An empirical result is considered robust if the pattern in question occurs across every single one of the 25 validation models. For example, "boyish" is masculine across all 25 models, and in every model is less masculine than "tough guy," since the full range of values for "boyish" is lower than the full range of values for "tough guy." 

\begin{figure}
\label{figure:3}
\caption{Gendered Meanings of Test Words}
\begin{center}
\includegraphics[scale=2.75]{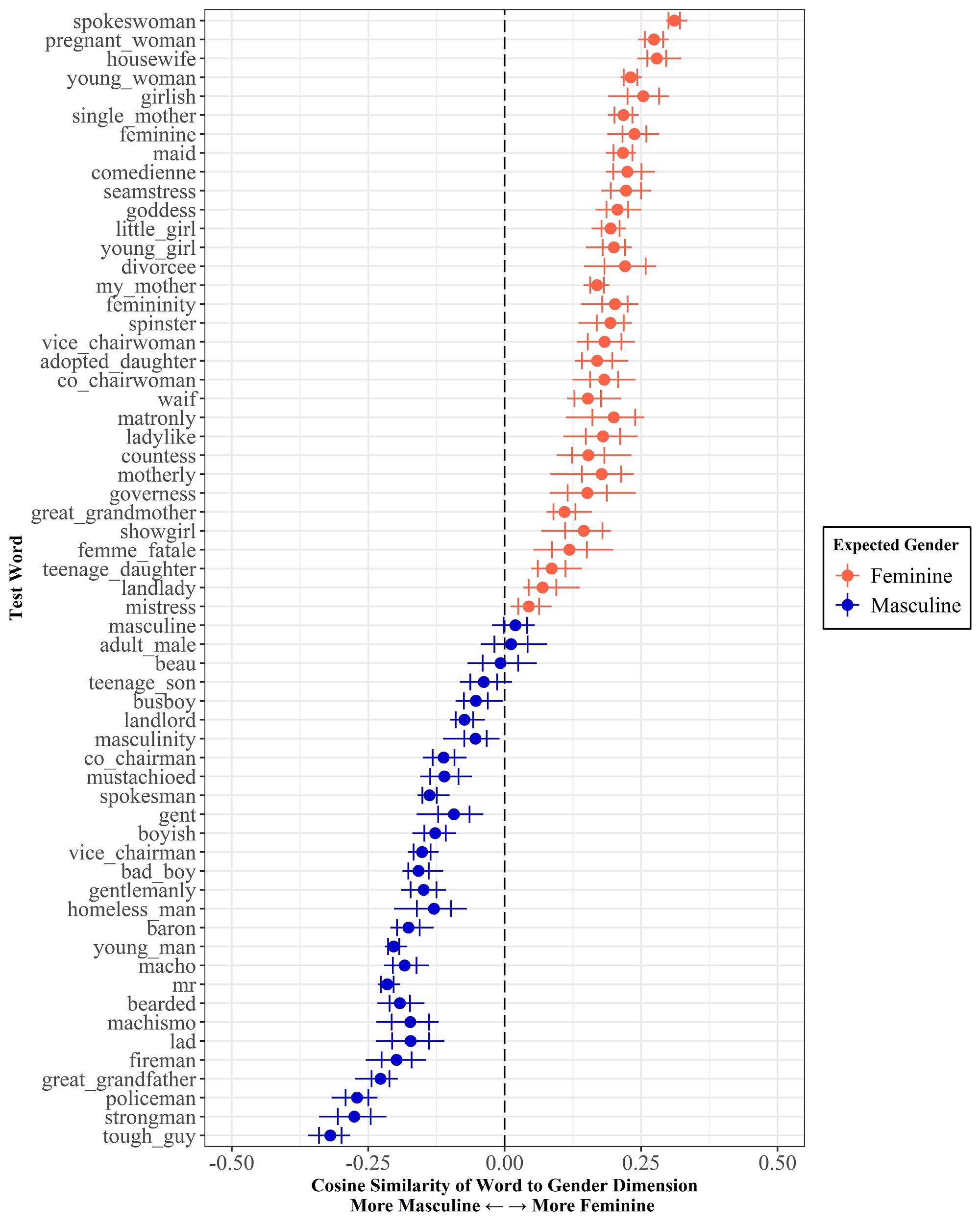}
\end{center}
\end{figure}

Next, we examined how our keywords about obesity connote gender (\href{figure:4}{Figure 4}). Consistent with literature reviewed earlier, we found that nearly all words describing obesity were feminine ("obese", "obesity," "morbidly obese," "overweight," "obesity epidemic", "flab," "gained weight," and "excess weight"). Indeed, our model learns to connote "obese" with women almost as strongly as it learns to "slender," and "slim" with women. The two exceptions include "flabby" and "being overweight" which do not consistently connote men or women across all 25 models, but on average connote women. Further, terms about slenderness connoted femininity. Most body descriptors about muscularity were masculine ("muscular," "burly," and "lean") while "toned," was feminine.

Terms about eating disorders, and especially "anorexic," were the most feminine terms. This result is consistent with previous literature, which showed that news articles about overweight tends to discuss women more often than men, and this pattern is exaggerated among articles about eating disorders \citep{saguy2010morality}. Interestingly, we found that most of our obesity-related terms were feminine, including terms about behaviors related to health and weight management, such as "diet," "dieting," "overeating," "healthy," "exercise," "sedentary," and "unhealthy." Our results suggest that not only do discussions of obesity generally occur in the context of women; discussions of the body, health, and weight management are in that context as well.

\begin{figure}
\label{figure:4}
\caption{Gendering of Obesity-Related Words}
\begin{center}
\includegraphics[scale=2.75]{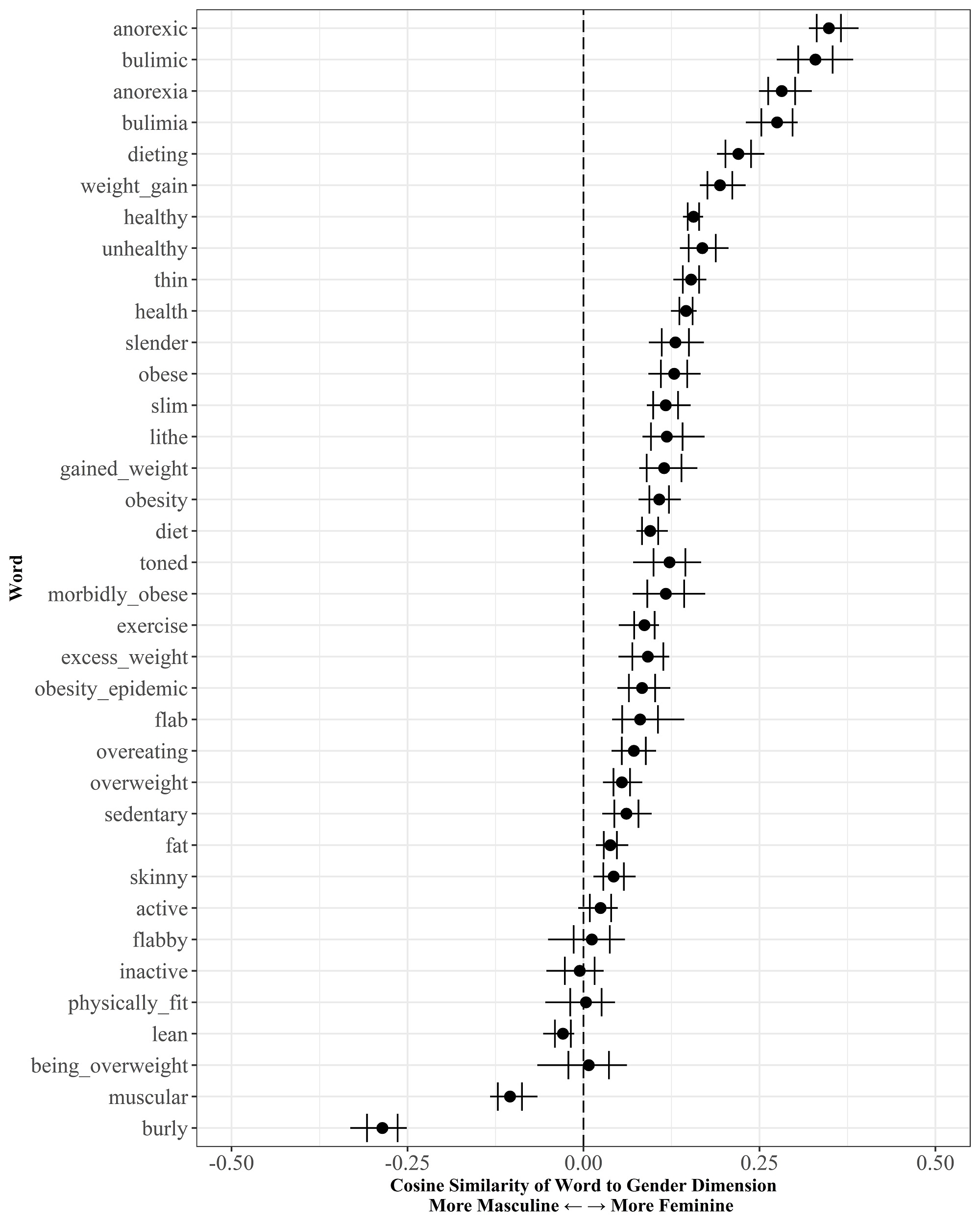}
\end{center}
\end{figure}

\section*{Immorality and Obesity} 
Our second dimension captures the concept of morality, and ranges from immoral to moral. On our model trained on all data, 60 (94\%) of 64 training words and 28 (97\%) of 30 testing words were correctly classified as immoral or moral. Across the validation models, this dimension robustly classified 56 (88\%) the 64 training words and 25 (83\%) of the 30 testing words. For example, "pristine" and "chastity" were classified as moral, while "profane," and "obscene" were classified as immoral (\href{figure:5}{Figure 5}). 

\begin{figure}
\label{figure:5}
\caption{Morality of Test Words}
\begin{center}
\includegraphics[scale=2.75]{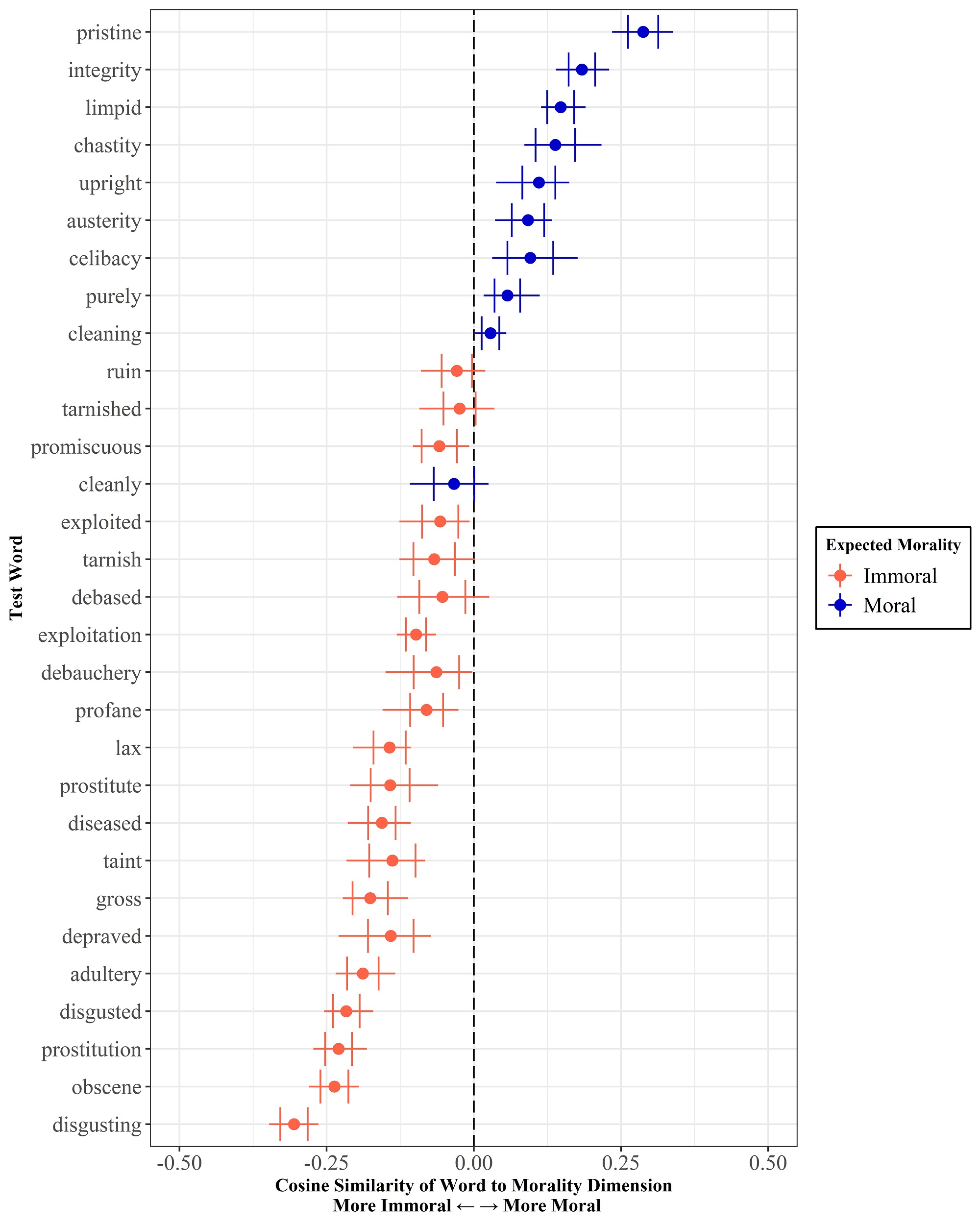}
\end{center}
\end{figure}

With respect to obesity keywords, words about obesity (e.g., "obese," "obesity," and "overweight") all connote immorality (\href{figure:6}{Figure 6}). While some words about slenderness connote morality (e.g., "slender," "slim," "lithe"), we find that "skinny," and "thin," tend to connote immorality. More generally, we see that many words about healthiness and healthy behaviors related to weight management connote morality (e.g., "healthy") and words indicating a lack of health (e.g., "unhealthy") tend to connote immorality. This is consistent with literature on the moralization of health and illness reviewed earlier. Indeed, we find that our health and morality dimensions substantially overlap in meaning, evident by their cosine similarity of .55 (possible range -1 to 1). Similarly, qualitative and historical work reviewed earlier suggests that moral and health meanings of obesity can be mutually-reinforcing. 

Previous literature finds that individuals with eating disorders are portrayed as victims while overweight individuals are portrayed as culpable for their weight \citep{saguy2010morality}. We find that not only are overweight and obesity classified as immoral, but that eating disorders are classified as immoral as well. Indeed, the second most immoral term in our list is "bulimia," after "being overweight." Examining the terms with the highest cosine similarities to terms about eating disorders, like those listed in \href{table:2}{Table 2}, sheds more light on these immoral undertones. For example, "self induced" is one of the most similar terms to bulimic, while "addict" and "alcoholic," are some of the most similar words to "anorexic."

\begin{figure}
\label{figure:6}
\caption{Morality of Obesity-Related Words}
\begin{center}
\includegraphics[scale=2.75]{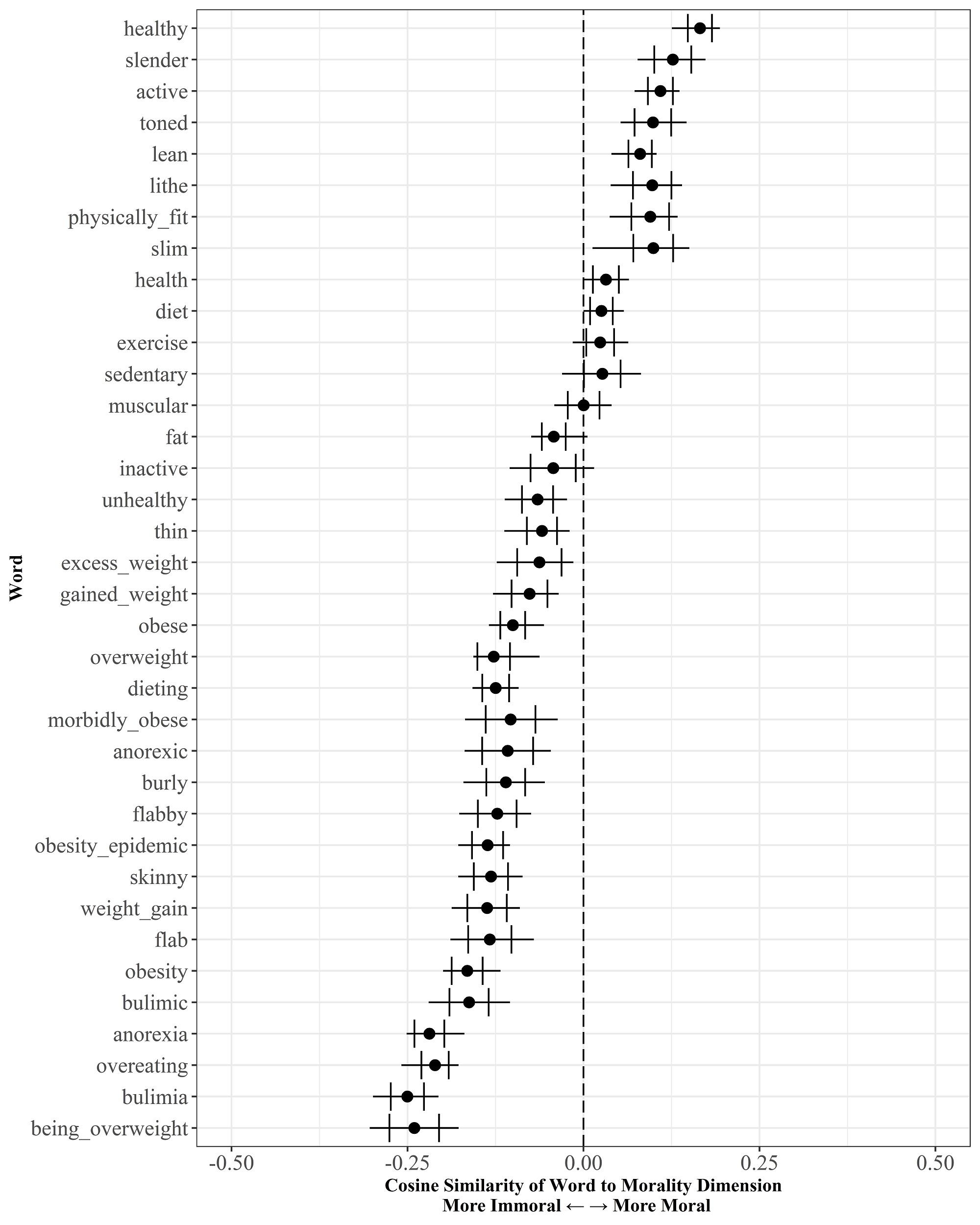}
\end{center}
\end{figure}

\section*{Health and Obesity} 

Our third dimension captures health, and ranges from unhealthy to healthy. On our model trained on all data, 121 (93\%) of 130 training words were correctly classified and 55 (92\%) of 60 testing words. Across the 25 validation models, this dimension robustly classified 111 (85\%) of the 130 training words. It robustly classified 52 (87\%) of the 60 testing words across all validation models. (\href{figure:7}{Figure 7}). For example, "wholesome," and "healthy habits" were classified as healthy, and "disease" and "poisonous" as unhealthy.\footnote{No words describing body weight were used in training or testing. We removed words form our obesity-lexicon that overlapped with the health anchor words.} 
 
\begin{figure}
\label{figure:7}
\caption{Health of Test Words}
\begin{center}
\includegraphics[scale=2.75]{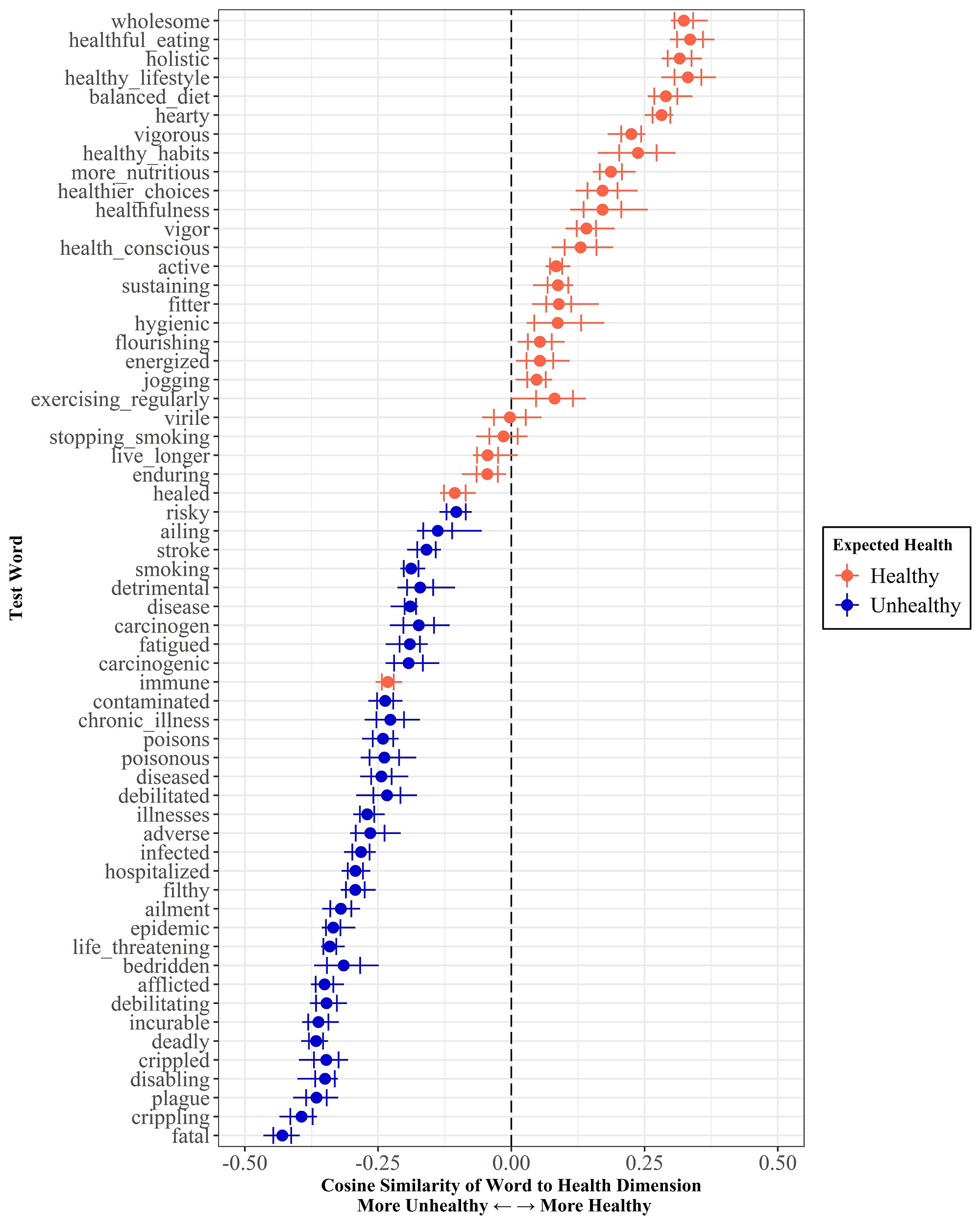}
\end{center}
\end{figure}

Next, we examined how our key terms about obesity connote health and illness (\href{figure:8}{Figure 8}). We find that all terms about fatness connote illness, consistent with literature reviewed earlier. While many terms about slenderness connote health, we find that "thin" and "skinny" are again outliers, similar to our findings for morality. "Skinny" is consistently unhealthy; "thin" is not consistently categorized, but is unhealthy in most models. These two terms denote the lack of fatness, but perhaps they also connote a lack of fatness due to underlying disease or malnourishment. We also find that eating disorders are all classified as unhealthy. In fact, "obese," "overweight," and "morbidly obese" carry connotations of unhealthiness that are approximately as strong as "bulimic," "bulimia," "anorexic," and "anorexia."

\begin{figure}
\label{figure:8}
\caption{Health of Obesity-Related Words}
\begin{center}
\includegraphics[scale=2.75]{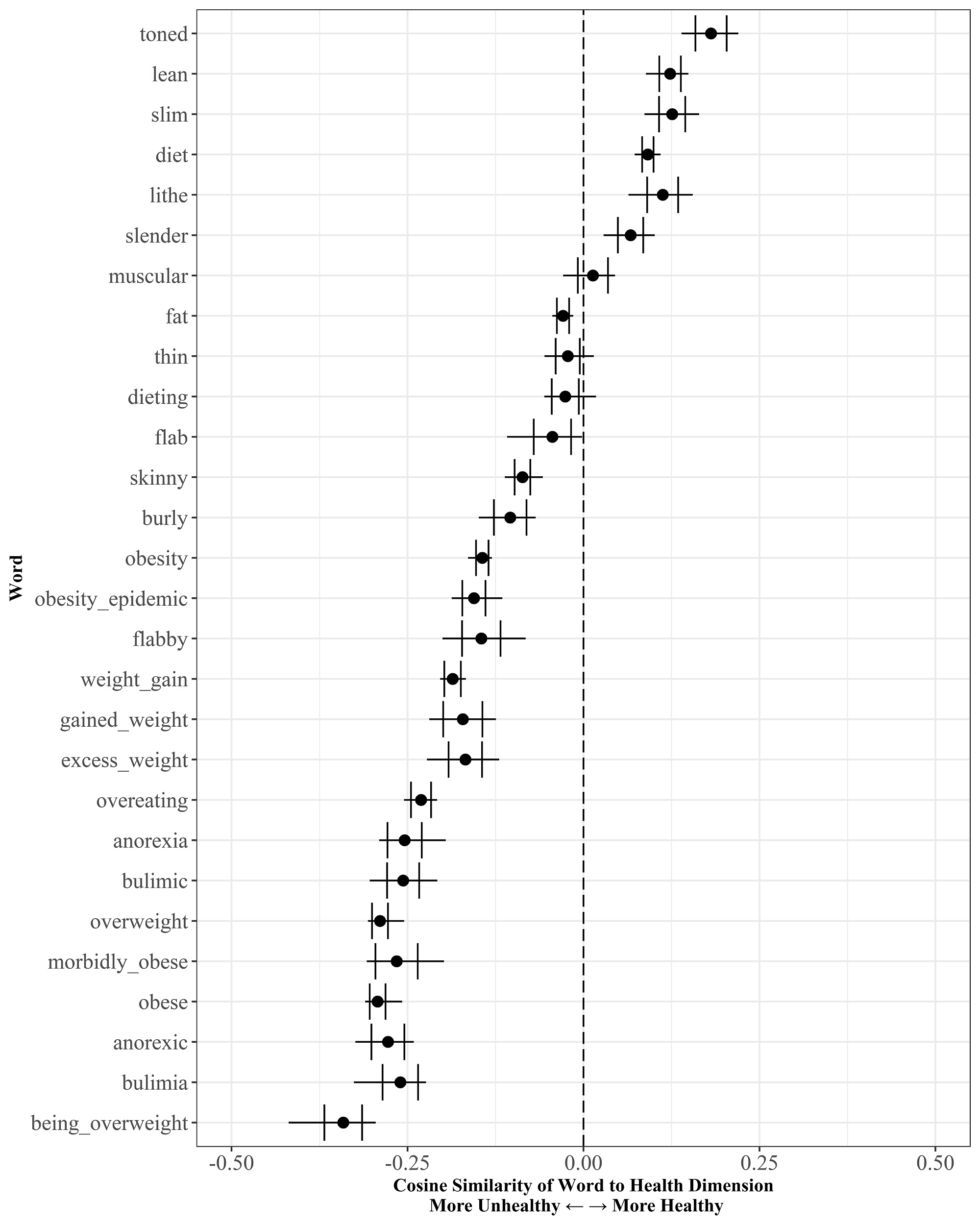}
\end{center}
\end{figure}

\section*{Socioeconomic Status and Obesity} 

Our fourth and final dimension captures Socioeconomic Status (SES), ranging from low SES to high SES (Figure 9). On our model trained on all data, 95 (95\%) of 100 training words were correctly classified and 59 (98\%) of 60 testing words. Across the 25 validation models, this dimension robustly classified 86 (86\%) of the 100 training words, and 56 (56\%) of the 60 testing words. For example, "multimillionaire," and "opulent" were classified as high SES, and "less affluent" and "indigent" as low SES (\href{figure:9}{Figure 9}).

\begin{figure}
\label{figure:9}
\caption{Socioeconomic Status (SES) of Test Words}
\begin{center}
\includegraphics[scale=2.75]{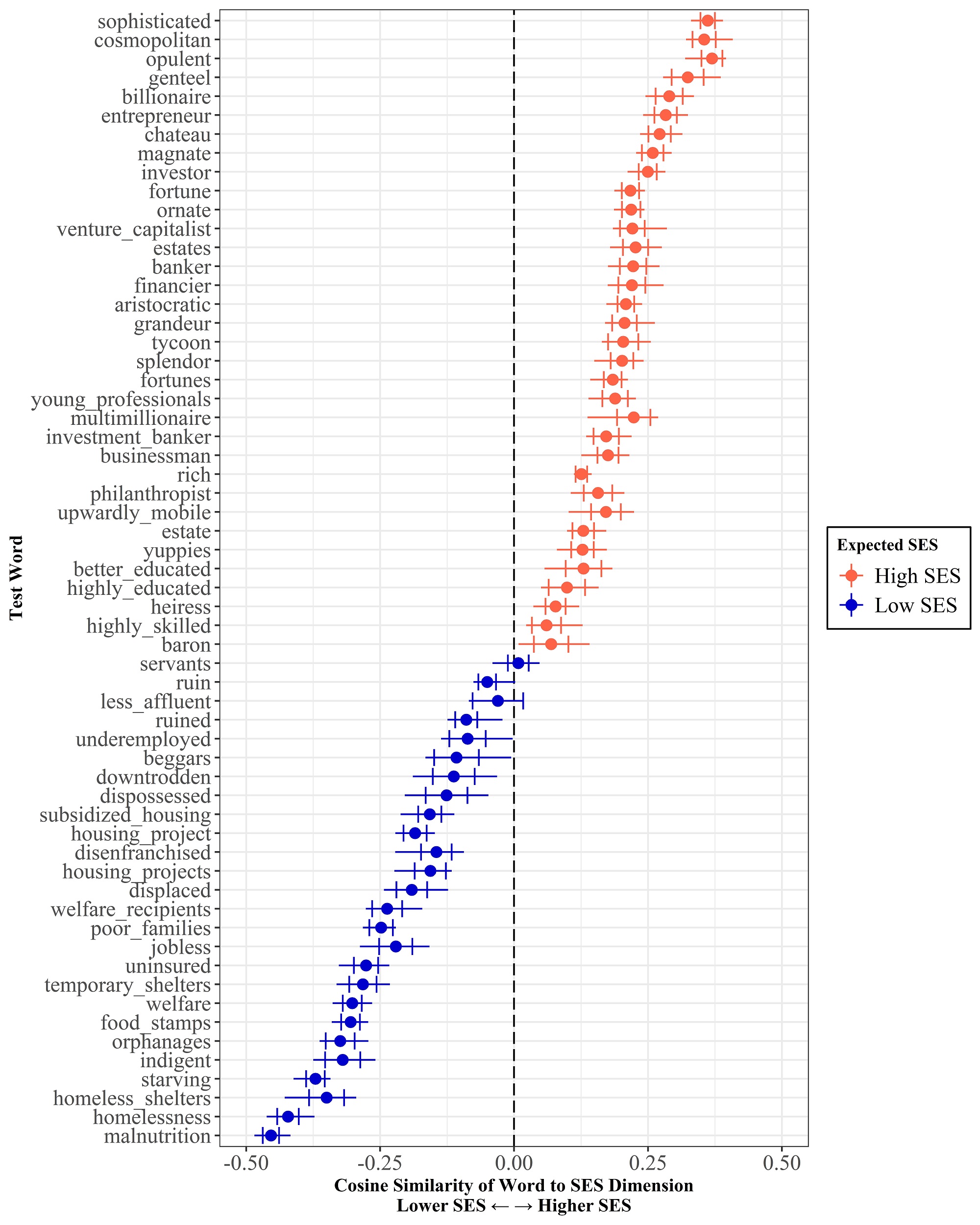}
\end{center}
\end{figure}

We find that nearly all terms about fatness connote low socioeconomic status: just "flab" and "fat," are inconsistently categorized. We also find that most terms about slenderness connote high SES, with the exception of "thin" which is again inconsistently categorized (\href{figure:10}{Figure 10}). 

Previous literature suggests that obesity is framed as a condition associated with low SES while anorexia is associated with high SES \citep{saguy2010morality}. However, we find eating disorders, in addition to "obese," "overweight," and "obesity," are associated with low SES; indeed, "bulimia" carries some of the strongest meanings of low SES out of all our key terms. More generally, we find that most weight and health related words are associated with low SES, including words such as "diet," and "overeating." As health issues are tightly intertwined with finance and social class, it is plausible that discussions of health at large are in the context of their financial burden. 

Finally, we find that morality and SES overlap substantially in meaning, corroborating qualitative literature reviewed earlier. The cosine similarity of these two dimensions is .30 (possible range of -1 to 1). This cosine similarity is lower than that between morality and health, suggesting that meanings of morality and SES are mutually reinforcing, but not as strongly as health and morality.

\begin{figure}
\label{figure:10}
\caption{Socioeconomic Status (SES) of Obesity-Related Words}
\begin{center}
\includegraphics[scale=2.75]{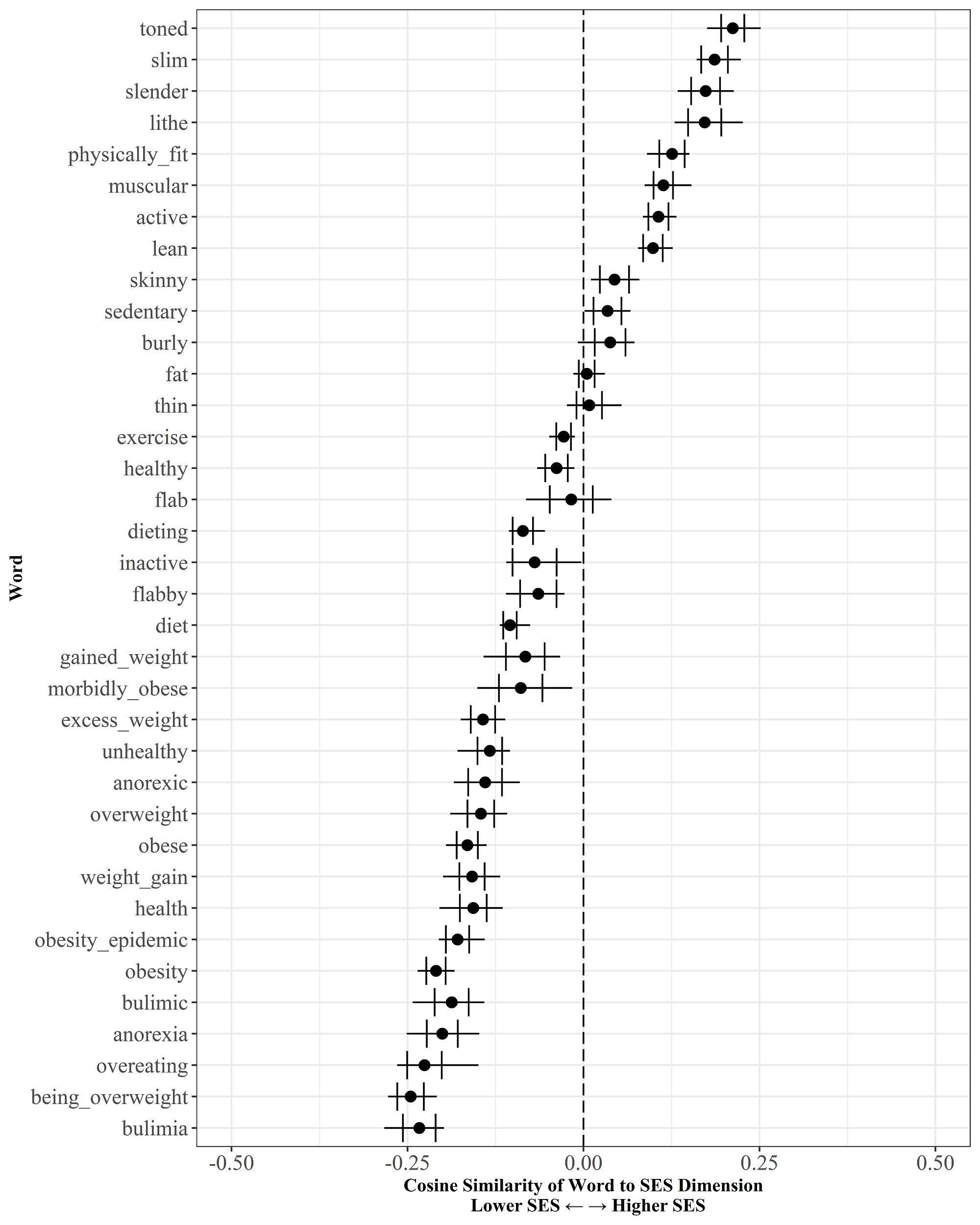}
\end{center}
\end{figure}

\vspace{5mm}
\begin{center}
\item{\section*{DISCUSSION}}
\end{center}
This paper provides a computational account of the learning of private culture (schemata) from public culture. We illustrate this approach by showing how schemata related to obesity can be learned from news discourse, using a new tool to model language: neural word embeddings. We suggest that neural embedding methods are not only computational tools, but also cognitively plausible models for schematic processing. Like humans, these methods aim to learn abstract, compressed representations from stimuli in their cultural environment. They represent and process semantic information as dense, distributed representations, and learn these representations by predictively coding stimuli. 

We model two types of schemata learned by word2vec: schemata of words (focusing on those related to obesity) and schemata of higher order constructs of gender, morality, health and social class. In addition to replicating prior findings on the pejorative meanings of obesity, we also find that even specific words used to refer to body weight - such as "thin" vs "slender," evoke subtly different associations. We validated our empirical findings with two alternative approaches to capture schemata and multiple other robustness checks. Neural word embeddings learn from large-scale text data; thus we suggest that training neural word embeddings on news data provides a concrete computational model for learning one kind of private culture (schemata related to words) from public culture -- and that this model is cognitively plausible at the computational, representation/algorithm, and hardware Marr levels. Embeddings begin with random representations of words (unlike humans); that said, this naivet\'{e} makes them useful tools to model the meanings that are activated and encoded from specific cultural environments, such as news. 

Our paper makes several theoretical contributions to the field of culture and cognition. While the distinction between private and public culture has garnered much attention \citep{sperber1996explaining,strauss1997cognitive,lizardo2017improving}, we provide new insights into how information travels between these two levels of culture \citep{foster2018culture}. Our model proposes schemata as a key ingredient in this process. Further, while schemata are considered fundamental to cultural sociology, there remains little consensus on what exactly a schema is \citep{wood2018schemas}. We have clarified the distinction between schemata and frames, and provided a concrete model for what a schema is and how it works. Most crucially, we suggest that schemata take the form of distributed representations, and that they can be learned online through predictive coding. The representations produced in this way encode higher-order structures, like the gender, morality, health, and SES dimensions, which can be recovered from the neural embeddings and used to unpack facets of their meaning.

Our paper also makes several contributions to computational sociology. We use embedding methods, which are increasingly being used in sociology and the social sciences for empirical analyses of culture (e.g., \cite{kozlowski2019geometry, jones2019stereotypical, stoltz2019concept}). Our paper theoretically motivates these methods by drawing parallels to culture and cognition, using Marr's levels of analysis \citep{foster2018culture, marr1982vision}, and proposing that embeddings model schemata (specifically, those having to do with semantic memory and word meaning). Along the way, we provide an extensive account of these methods and how to implement them for robust empirical research. For instance, we illustrate how to choose modeling hyperparameters, how to robustly measure higher order constructs in embeddings, and how to compare the quality of overall models against two standard benchmarks in computer science (Google Analogy test and the WordSim test). To address concerns around the instability of estimated embedding models \citep{antoniak2018evaluating}, we also present empirical results based on aggregating results from multiple embedding models trained on our data. We present two alternate methods to extract cultural dimensions, and use classification tasks to test the quality of extracted dimensions. Finally, we provide accompanying tutorials on GitHub so that other scholars may reproduce our methods and use these methods in their own research. 

We also make several contributions to the sociology of gender, the body, and health. While a large body of scholarship documents public and private meanings of body weight, and implicates media consumption as a core influence on our meanings of body weight, this paper illustrates exactly how meanings of body weight may be learned and reinforced from media exposure. Consistent with qualitative literature reviewed earlier, our models learns to associate heavy body weight with women, immorality, unhealthiness, and low class. Our empirical results also suggest that obesity may be gendered not only because it is about \textit{heavy} body weight, but just because it is about health and body weight---and health and body weight more broadly connote femininity. For example, we find that words like "diet," "health," and "overeating" are connoted with women as well. These findings make sense given the gendered history of weight control \citep{stearns2002fat}. Our empirical results also show the constellation of words activated by key terms about body weight (\href{table:2}{Table 2}) in news. These illustrate more inductively, how these terms are charged with evaluative meanings well beyond body weight and health. For example, "physically fit," activates meanings similar to "open minded," "competent," and "personable," while "bulimic" activates meanings similar to "irritable," and "slim" activates meanings similar to "blond." Similarly, a prior study found that weight bias (measured implicitly or explicitly) powerfully influences perceptions of both the behavior and character of obese people \citep{schwartz2006influence}. Further, while previous literature on health, obesity, and social status often discusses how the meanings around obesity are mutually reinforcing we empirically show that these meanings are learned in parallel and quantify the extent to which they are mutually reinforcing.  

Additionally, our empirical results highlight how seemingly similar terms regarding body weight evoke subtly distinct meanings. For example, "slender," connoted health, morality, and high-class, as we expected from qualitative work. In contrast, "thin," was learned as immoral and was not consistently learned as low-class or high-class, or healthy or unhealthy. While previous literature on culture and the body often refers to various terms for slenderness interchangeably, as if both symbolize high status, health, and virtue, our results suggests that words activate distinct meanings. Similarly, prior work has found that terms for heavy body weight evoke subtly different meanings \citep{vartanian2010obese}. Neural word embedding methods offer a way to capture such subtle, but systematic patterns in meaning, and do so at scale. 

\vspace{2mm}
\subsection*{Limitations and Directions for Future Research}

Our work also has several limitations. For instance, while we replicate meanings of obesity which are extensively documented in prior work, validate our empirical findings in numerous ways, and illustrate that our model can solve linguistic tasks in ways that correspond to humans solutions to these tasks (e.g., the Google Analogy Test and WordSim 353 Test), we do not compare word-level schemata around obesity learned by our model with extrinsic cognitive data around obesity terms. However, an abundance of previous work has already established that humans' private meanings \textit{substantively} correspond to semantic information in embeddings and distributional models more broadly  (e.g., \cite{caliskan2017semantics, bolukbasi2016quantifying, kozlowski2019geometry, grand2018semantic, jones2019stereotypical, joseph2020word, garg2018word, dehghani2017decoding, vodrahalli2018mapping}). Our paper builds on this established finding, to show that neural embeddings also \textit{learn} these meanings in a cognitively plausible way. 

Further, our work is limited in that we investigate only two types of structures for schemata: word-level schemata and single dimensions. As noted by \cite{kozlowski2019geometry}, dimensions operationalize binary opposition - a core structure of meaning in structuralism but certainly not the only possible structure \citep{levi1963structural}. In a binary opposition, meaning emerges from the opposition of two poles, such as the concept of gender being defined by the opposition between masculinity and femininity. Other schemata, such as schemata around race and sexual orientation, might be better captured with higher order structures such as subspaces carved out by sets of dimensions. Future work, therefore, might not only investigate other meanings but also other possible structures of meaning using embeddings, or assumptions about meaning structures themselves. For example, L\'{e}vi-Strauss suggested that binary oppositions were not only prevalent in public culture, but also a core structure in private culture. Future work might investigate the extent to which binary oppositional structure comes from the patterns in language, from the way we measure dimensions in embeddings, or from the way that humans process meaning.

Our modeling is also limited in how it accounts for context. Our embedding model learns a single vector to represent the meaning of each word; thus each vector smooths over the many reconstructions of its word.\footnote{As we illustrated in our results, for example, polysemous words like "fat" are given a single representation despite the fact that "fat" occurs in systematically distinct contexts -- sometimes referring to a body and other times to a macronutrient} However, humans likely reconstruct a word's meanings each time we encounter the word, even if there is some degree of stability in the way that meanings are reconstructed across encounters. Future work might use contextualized neural word embeddings \citep{devlin2018bert,peters2018deep}, which model each instance of a word as a vector, rather than modeling all instances of that word in a corpus with a single vector. Ideally, modeling should account for both stability and context sensitivity in schemata \citep{smith2009distributed}. 



Our paper has focused on learning \textit{shared} private meanings from a very specific cultural diet (\textit{New York Times} articles on obesity). However, humans hold a wide array of private meanings and these are shared to varying extents. Heterogeneity may come from the learning process (indeed neural word embeddings are not deterministic learners), but also from variety in the public cultural environment. Among our many trained models (validation models trained on the New York Times, and a published model trained on GoogleNews), while all models learn to connote heavy body weight with women, immorality, illness, and low class, they do so to varying degrees. More realistically, humans have a very diverse cultural diet which can lead to greater heterogeneity in private meanings. Future work might model heterogeneity in meanings learned when neural embeddings are exposed to varying cultural diets, or exposed to the same cultural diet in different ways (i.e., curriculum training \citep{bengio2009curriculum}). 

Finally, while our model learns from text data, the cultural environment includes a variety of modalities such as television, advertisements, songs, and radio. In fact, our theoretical arguments generalize beyond word2vec, and beyond text data \citep{gunther2019vector}. A variety of neural network architectures and distributed representations are ubiquitously used to process text (e.g., \cite{young2018recent, goldberg2016primer}), images (e.g., \cite{kriegeskorte2015deep}) and sound (e.g., speech) \citep{liu2017survey}. Some models may even simultaneously learn from multiple modalities.\footnote{Multimodal learning could reflect that certain private meanings may be more grounded in sensory experiences (i.e., embodied) \citep{lakoff1980conceptual, roy2005grounding, lucy2017distributional, baroni2016grounding, bruni2014multimodal,gunther2019vector}. Previous work has suggested that while concrete concepts are more likely learned from sensorimotor experience, abstract concepts like "freedom" cannot be learned through such experience. Distributional models provide a mechanism (language) for learning these more abstract concepts \citep{borghi2017challenge, gunther2019vector}.} Many of these neural-network-based models are also often predictive and rely on the distributional hypothesis - that meaning emerges from context. Future work could consider these wider varieties of models and modalities for additional empirical and theoretical insights into schematic processing.

\section*{Implications} 

This paper responds to calls to integrate computational methods and theorizing about cultural phenomena \citep{mohr2013introduction, wagner2015ontologies,foster2018culture, edelmann2020computational}. We not only use a new method (word embeddings) as a tool to empirically investigate cultural meanings, but also theorize that this tool is a model for learning private culture (schemata) from public culture. Using Marr's levels of analysis, we then illustrate the extent to which our method is a cognitively plausible account of schematic processing. Similarly, Marr's levels of analysis may be used to integrate other cultural theories with computational methods. Given the abundance of new methods in machine-learning this approach will be useful to integrate these methods with cultural theory. 

Theorizing a new method, like embeddings, also helps us to understand the method itself, and our implicit assumptions when using this method as a tool. In our paper we provide a detailed exposition of word2vec and how to use it robustly, as well as an accompanying tutorial. Our strategies to do robust empirical work with embeddings generalize beyond word2vec, and to other methods for estimating embeddings such as GloVe and FastText \citep{pennington2014glove, bojanowski2017enriching}. We hope these tools and strategies will enable other researchers to investigate a wide range of cultural phenomena. 

Machine-learned models are not merely methods for researchers but also objects pervasive in our environment. Models, like neural word embeddings trained on GoogleNews, structure many of our everyday experiences - from the recommendations, summaries, and search results we receive, to the text we read that has been automatically translated. Our work validates widespread concerns that machine-learned models, such as word2vec, encode pejorative meanings as they learn (e.g., \cite{bolukbasi2016quantifying, caliskan2017semantics}). Word2vec, like other machine-learning models, is not inherently biased against any specific people or attributes. However, in learning from human data they learn socially unacceptable (but pervasive) meanings such as gender biases. Thus, when they are used used in downstream tasks, these models may reproduce and even amplify our pejorative schemata. Many interventions into machine-learned bias focus on biases against legally protected attributes (e.g., \cite{manzini2019black, bolukbasi2016quantifying, zhang2018mitigating}). Weight discrimination remains legally unprotected in most states, and even socially acceptable, despite rising in prevalence \citep{charlesworth2019patterns, andreyeva2008changes}. Our paper highlights that machine-learning bias research needs to consider a broader array of stigmatizing attributes, such as obesity.  

Schemata are not inherently harmful; they are merely mental structures that enable us to efficiently interpret social life. However, they become harmful when they systematically and unduly constrain groups of people's well-being and life chances. Then, schemata (e.g., that obese individuals are lazy or immoral) reinforce stigma, discrimination, and, ultimately, inequality. Given the pervasive and systematic nature in which our words are loaded with negative meaning about body weight, it may be unsurprising that Americans hold, and act on, deeply internalized negative beliefs of body weight, in the forms of body dissatisfaction, eating disorders, fat prejudice, and discrimination on the basis of body weight. Similarly, our words may be vehicles to transmit and reinforce many other stigmatizing schemata, such as those around mental illness, sexual orientation, and race.

\vspace{5mm}

\section*{Acknowledgments}
This paper is based on work supported by the National Science Foundation Graduate Research Fellowship Program under Grant No. DGE-1650604, the Graduate Research Mentorship fellowship at UCLA, and the Graduate Summer Research Mentorship fellowship at UCLA. We are grateful to Abigail Saguy for her feedback on early drafts of this paper, and to Hongjing Lu, Andrei Boutyline, Omar Lizardo, and students in the graduate course on "Cognitive Sociology of Culture" at the University of Michigan for their feedback on more recent versions. We also thank audiences at our presentations at the American Sociological Association Meeting, Text Across Domains (TextXD) Conference, and the UCLA Computational Sociology Working Group for their feedback.

\bibliography{exportlist_6.1.18}
\bibliographystyle{asr}

\vspace{5mm}
\appendix
\newcommand*{\Appendixautorefname}{appendix}
\section{Vectors and Cosine Similarity}\label{appA}
Most simply, a vector is a list of $N$ numbers. These N numbers indicate a position in an $N$-dimensional vector space. For example, consider the familiar Euclidean plane; positions in this plane are indicated in Cartesian coordinates with reference to the $x$ axis and the $y$ axis. In fact, any location on this plane is described with a vector with 2 dimensions, the first number of which indicates its position along the $x$ axis and the second of which indicates the position along the $y$ axis. These two axes may also be thought of as (multiples of) two vectors, the $x$ axis as (1,0) and the $y$ axis as (0,1). Then, any point on the plane can be written as weighted combinations of these two vectors. The point’s position along the $x$ axis is some multiple of (1,0) and its position along the $y$ axis is some multiple of (0,1). Since any point on the plane may be written as a unique linear combination of these two vectors, they are said to provide a complete \textit{basis} for the Euclidean plane.

Mathematically, the cosine similarity of two vectors is proportional to their inner product (dot product), as illustrated below. Conceptually, this is a measure of the angle between two vectors (specifically, the cosine of that angle). Two vectors pointing in the same direction have a cosine similarity of 1; if they are pointing in exactly opposite directions, their cosine similarity is -1; and if two vectors pointing 90 degrees apart they have a cosine similarity of 0, also called "orthogonal." 

\begin{equation}
\textrm{cosine similarity}(A,B) = \frac{\sum_{i=1}^{n}{A_i B_i}}{\sqrt{\sum_{i=1}^{n}{{A_i}^2}} \sqrt{\sum_{i=1}^{n}{{B_i}^2}}} 
\end{equation}
where $A$ and $B$ are two $N$-dimensional vectors, and $A_i$ is the component of $A$ in the $i^{th}$ dimension, while $B_i$ is component of $B$ in the $i^{th}$  dimension.

\section{Data Cleaning and Preprocessing}\label{appB}

We cleaned data from LexisNexis in several steps. After downloading our news data from LexisNexis, we observed that many articles focused on recipes. We excluded articles that included the word "recipe" or "recipes," eliminating 8,495 (8.2\%) of the collected articles. Data were then pre-processed, including normalization of letters to all lowercase, removal of punctuation, and separation into sentences. Sentences included any phrases that ended with a semi-colon, colon, exclamation point, double quotes, period, or question mark. We also removed vocabulary words occurring less than 40 times in the data to prevent word2vec from learning poor-quality word-vectors. This process yielded a vocabulary size of 81,250 words, and total of 92,785,779 tokens.

\section{Selecting Model Hyperparameters and The Google Analogy Test}\label{appC}

The Google Analogy Test includes a series of analogies divided up into 14 subsections, including world capitals, currencies, family, and a series of syntactic sections such as tense and plurality \citep{mikolov2013efficient}.  For example, an analogy in the semantic section on World Capitals will include: "Berlin: Germany as Paris: ?," and if the model responds correctly it will return: "France." An analogy in the semantic section on family will include: "Mother: Father as Daughter: Son." In the syntax section with tense, a sample analogy might be "Walk: Walked as Run: Ran."  The Google Analogy test includes approximately 20,000 questions. To assess performance, we prioritized sections about family and syntax, since our corpus is a US news source curated to focus on obesity discourse rather than international and political news.

We chose hyperparameters across various stages. First, we decided our learning architecture (Continuous Bag of Words (CBOW) or Skip-Gram) and decided among two possible algorithms that speed up learning (hierarchical softmax or negative sampling). These two parameter choices, each with two options, yielded four candidate models. For all models, we found word-vectors of 500 dimensions; dimensionality usually ranges from 100—600 but the gains in performance diminish after 300 dimensions and models become more computationally expensive \citep{mikolov2013efficient}. We also selected a context window of 10 words (10 on either side of the target word). Context windows usually range from 2—10. We chose a larger context window as word-vectors developed on longer context windows tend to better capture the topical relationships between words, while shorter context windows tend to better capture syntactic and functional similarities; the former is the priority in this study \citep{goldberg2016primer,levy2014dependency}. For a detailed description of word2vec hyperparameters, see \cite{rong2014word2vec}. Our two hyperparameter decisions, each with two options, yielded four candidate models. 

The candidate model which finds vectors with CBOW and negative sampling (Model A) stood out among the four candidate models. We used the default number of 5 negatives samples. We then tested if this model performed better on the Google Analogy Test at lower dimensions (50-400), or at a smaller context window size of 5, but did not find a model that performed better. For all other hyperparameters, we used the default settings in the Gensim implementation of word2vec \citep{rehurek_lrec}. See Table 1 for detailed performance results. The hyperparameters of Model A are used in all subsequent word2vec models in this paper. 

\section{Developing Training and Testing Sets of Words}\label{appD}
To decide on words to use to develop our scales for gender, health and SES, we iteratively followed these steps:

\begin{enumerate}

\item{Brainstorm an initial list of relevant anchor words (e.g., wealthy and poor). Do not include any words about body weight or shape. For gender words, we used words from \cite{bolukbasi2016quantifying} as an initial list.}
\item{Check for completeness by using a thesaurus and examining the most similar words to brainstormed words in the word2vec model.}
\item{Separate out a set of training words to evaluate scale performance.}
\item{Compute the training accuracy on these training words to see how well words are dichotomized.}
\item{If training words are poorly dichotomized, revise training words. Some words are rare, or hold multiple meanings and so may not be good training words. For example, the words with the highest cosine similarity to "rich" in our training model were related to taste and texture, not affluence, and so we excluded "rich" as a training word.}
\item{Perform cross validation on training words (e.g., leaving one word-pair out in each fold).}

\end{enumerate}

For each dimension, we used equal numbers of anchor words to represent each end of the dimension. For gender, health and SES, words were also paired for the Bolukbasi method, which requires paired words. As described earlier, for a list of words to represent morality, we used lexicon words in the purity/impurity dimension from Moral Foundations Theory. Sample training and testing words for each dimension are listed in \href{table:4}{Table 4}. For a complete list of training and testing words, see our \href{https://github.com/arsena-k/Word2Vec-bias-extraction}{code on GitHub}. 

\begin{table}
\begin{center}
\caption{Sample Training and Testing Words}
\label{table:4}
\begin{tabular} { p{2.5cm} p{5.25cm} p{5.25cm} }
 \noalign{\vskip 1.7mm}  
   & \textbf{Sample Training Words} & \textbf{Sample Testing Words} \\
 \noalign{\vskip 2.5mm}  
   \multicolumn{3}{l}{\textbf{Gender}} \\
 \hline
 \textit{Feminine} & hers, herself, she, lady, gal, gals, madame, sisterhood, girlhood, matriarch  & goddess, girlish, feminine, ladylike, mistress, landlady, seamstress, waif, femme fatale, comedienne \\
 & & \\
 \textit{Masculine} & his, himself, he, guy, dude, dudes, sir, brotherhood, boyhood, patriarch & boyish, masculine, lad, policeman, macho, mailman, bearded, mustachioed, gent, strongman \\
  \noalign{\vskip 2mm}  

  \multicolumn{3}{l}{\textbf{Morality}} \\ 
  \hline
 \textit{Moral} & modestly, sacred, saintly, sterility, holy, virtuous, saint, decency, decent, piety &  austerity, chastity, cleaning, integrity, purely, pristine,  celibacy, upright, limpid, cleanly \\ 
  & & \\
 \textit{Immoral} & sin, sick, stains, contagious, blemish, filthy, trashy, sinners, lewd, repulsive &  taint, obscene, prostitution, gross, profane, tarnish,  debased, disgusting, lax, adultery \\ 
  \noalign{\vskip 2mm}  

  \multicolumn{3}{l}{\textbf{Health}} \\
 \hline
 \textit{Healthy} & fertile, healthy, healthy, safer, salubrious, strength, sanitary, clean, healing, heal &  fit, flourishing, sustaining, hygienic, hearty, enduring, wholesome, holistic, healed, immune \\
  & & \\
 \textit{Unhealthy} & infertile, ill, potentially harmful, riskier, deleterious, weakness, filthy, dirty, harming, hurt  & deadly, diseased, adverse, crippling, carcinogenic, incurable, disease, stroke, plague, smoking \\ 
  \noalign{\vskip 2mm}  
  
  \multicolumn{3}{l}{\textbf{Socioeconomic (SES)}} \\
  \hline
 \textit{High SES} & wealth, wealthier, wealthiest, disposable income, wealthy, suburban, upscale, abundant, white collar, aristocrat & rich, billionaire, banker, fortune, heiress,  financier, tycoon, baron, grandeur, genteel \\ 
  & & \\
 \textit{Low SES} & poverty, poorer, poorest, broke, needy, slum, slums, lacking, blue collar, peasant & downtrodden, homeless shelters, indigent, jobless, welfare, temporary shelters, dispossessed, welfare recipients, food stamps\\ 
 \hline
 \multicolumn{3}{m{14cm}}{ Notes: Some training words were used multiple times, but matched up in different word-pairs, to increase the training sample size. Testing words were all unique. } \\

\end{tabular}
\end{center}
\end{table}

\section{Robustness Checks}\label{appE}

We extracted four dimensions (gender, morality, SES, and health) in our word2vec models, using the approach described in our methods, in which we averaged anchor words at each end of the dimension and took the difference between those averaged vectors. We checked the robustness of our findings with respect to: 1) accuracy of the method at classifying words known to lie at one end of the dimension or the other 2) sampling of the data used to train word2vec models, 3) selection of anchor words to extract the dimension, and 4) the method used to extract the dimension. Our extensive validation is partly motivated by larger concerns that empirical results from word embedding models may not be stable across corpora used for training models, and can even be sensitive to the presence of specific documents \citep{antoniak2018evaluating}. However, we find that our results are robust across all four of these facets. We describe (1) and (2) in the main paper, and describe (3) and (4) below. 

Our third validation check is on our word-selection. After settling on anchor words, we examined how variable performance was to our choice of anchor words. Specifically, we performed cross-validation on the training word-pairs (for a given dimension, splitting the anchor word-pairs into 10 parts, and the iteratively extracting a dimension on nine of these parts and measuring accuracy on the  held-out anchor word-pairs). Across all dimensions, mean accuracy across training subsets ranged .92 to .96, and mean accuracy on held-out subsets ranged .82 to .93. The high performance across various subsets of training words suggests that the extraction of dimensions is not sensitive to word-choice. As an additional check to ensure our results for gender and morality are not sensitive to our word choices, we also replicated our results using a more restricted, established lexicon for gender from \cite{bolukbasi2016quantifying}, and using the full lexicon from the Moral Foundations Theory to capture morality, rather than just terms from the purity dimension of this lexicon. 

Our fourth validation check is on our methods used to extract dimensions. We confirmed our empirical findings regarding obesity keywords using two alternative methods to extract dimensions and classify words: a second geometric method developed by \cite{bolukbasi2016quantifying} and a method using a machine-learning classifier. The Bolukbasi method uses Principal Components Analysis, rather than averaging, to reduce noise across the collection of anchor words. This method yielded almost the exact same dimensions as the one described in the main text. On our model trained on the full corpus, the cosine similarity between the gender direction extracted with the Bolukbasi method and the gender direction extracted with the Larsen method was .98. For health, and SES, this cosine similarity was .95 and .97, respectively. These high cosine similarities suggest that the same concept is found for each dimension regardless of our choice of the two extraction methods. Since this method requires equal numbers of anchor words for each pole of a dimension (e.g., equal numbers of masculine and feminine words), we do not examine our morality direction. Our morality direction uses an established but unbalanced lexicon. Using the Bolukbasi method on our model trained on all data, testing accuracy across gender, health, and SES dimensions ranged from 95\%-98\%. We found that classification of obesity-keywords largely corroborated empirical findings from the Larsen method. 

Our other alternative method included training four machine-learning classifiers to predict which end of each dimension (e.g., masculine or feminine) our training word-vectors were closest to. More technically, we use linear Support Vector Machines (SVM), which learn a hyperplane that discriminates the word-vectors at one end of the dimension and word-vectors at the other end of the dimension. This is a classification problem, since we had many word-vectors, each with $N$ dimensions, and want to predict a binary outcome. Machine-learning classifiers like SVM risk overfitting when the number of features greatly exceeds the number of training samples. We risked this since we had few training words (100-170 depending on the dimension) compared to the dimensionality of each word (500). Thus, for this robustness check we retrained a word2vec model on all data with 300-dimensional vectors, rather than 500 dimensions. For each of these four classifiers, we used cross validation across subsets of training words to select a value for the hyperparameter $C$ (strictness of the classification) and then tested our trained model on our testing words. Using the SVM method, testing accuracy across our four dimensions ranged from 90\%-98\%. See Table 3 for detailed accuracy results. 
We then used these trained SVM models to predict whether our obesity keywords lay at one end of each dimension or another. Initially, we found that these results corroborated empirical findings from both geometric methods regarding all dimensions except gender. This was surprising given our high training/testing performance for gender classifications. 

We speculated that this mismatch may be because gender training words are the most "sharply" or "widely" separated - our training words were too easy. Indeed, many of our training words for gender are explicitly and largely defined by their gendered meanings, such as "he", "him", "man" and "machismo." In contrast, training words for other dimensions, like "yuppie" for classifying SES, carry more layers of meaning than just social class. "Yuppie," for example, also carries meaning about age and urbanization. Thus, the gendered differences in our training set may be \textit{too easy} to classify, leading the model to overfit to these training cases. Equivalently, if we trained a classifier to differentiate "blues" and "greens" but only provided training examples of royal blue and forest green, the classifier will perform poorly if asked to differentiate sky blue, navy, olive, or lime green. To test for this, we added more implicitly gendered training words (e.g., "independent" as masculine, and "dependent" as feminine) and removed some of the explicitly gendered words. We used two versions of updated training words, to vary the number of word-vectors added and removed. In both cases, words about body weight were now classified with respect to gender in ways that corroborated findings from our other two methods. When we used these revised training sets for the other two classification methods, there were no changes in initial empirical findings. These results corroborate that the disadvantage of a machine-learning approach compared to the two geometric approaches could be its tendency to overfit. Indeed, the machine-learning approach requires many, diverse training examples to learn a pattern robustly. 

\end{document}